# Discrimination vs. Generation: The Machine Learning Dichotomy for Dopaminergic Hit Discovery

Temitope Sobodu[1,2], Adeshina Yusuf[3], Dan Kiel[2], Dong Kong[1]

[1]Department of Endocrinology, Boston Children's Hospital, Harvard Medical School, Boston, MA, USA.
[2]Department of Pharmacology, Massachusetts College of Pharmacy and Health Sciences, Boston, MA, USA.
[3]Center for Computational Biology, University of Kansas, Lawrence, Kansas, USA.
Email: Dong.Kong@childrens.harvard.edu, Temitope.sobodu@childrens.harvard.edu.

## Abstract

Virtual screening plays a pivotal role in early drug discovery, traditionally dominated by physics-based methods. While these approaches offer detailed insights, they are often hindered by high computational costs, limited sampling, and forcefield inaccuracies. Advances in Machine Learning (ML) and Deep Learning (DL) present resource-efficient alternatives, with approaches like predictive geometric ML and generative geometric ML showing promise in enhancing both efficiency and predictive capability. Here, we compare these two strategies, retrospectively and prospectively, for identifying novel agonists targeting the dopamine D2 receptor. To complement DIFFDOCK's dual functionality in protein-ligand conformer generation and confidence estimation, we adopted a complementary atom-type-based confidence model for EQUIBIND. This pipeline, termed the discriminative model, integrates a featurization step and an XGBoost classifier to differentiate between active and inactive ligands. The top-ranked compounds from both models were evaluated using an ultrafast dopaminergic biosensor assay, dLight. Our results demonstrate that the generative model achieved a higher hit rate, notably leading to the discovery of Compound 1, a nanomolar dopamine D2 receptor agonist with a novel scaffold.

The discovery of novel therapeutic agents is a cornerstone of drug development, and virtual screening has emerged as a vital tool in the early stages of this process. Traditional physics-based virtual screening methods, though robust, are often hindered by their substantial computational demands[1,2]. These approaches typically require extensive candidate sampling, followed by rigorous scoring and refinement steps, making them resource-intensive and time-consuming[2].

In response to these challenges, the field has increasingly turned to Machine Learning (ML) to streamline drug discovery[1,3]. ML offers the potential to predict molecular interactions with greater efficiency, thus reducing the computational burden associated with traditional methods[4,5]. Two promising ML strategies have gained prominence: predictive geometric deep learning and generative geometric deep learning[6,7]. Regression-based models, such as EQUIBIND[8], circumvent exhaustive sampling by directly predicting the final pose of a ligand. On the other hand, generative models, such DIFFDOCK[9], aim to generate near-native poses, offering a novel approach to hit discovery. Here, we compared both approaches in the context of a dopaminergic hit discovery task. The models employed consist of the following components.

- A discriminative model, which consists of a geometric graph neural network docking model — EQUIBIND[8] layered with an atomic interaction-based scoring function[10], and a classifier model that differentiates and active from an inactive dopamine receptor ligand, in combination we termed these layers the EFX pipeline.
- A generative model which deploys a diffusion-based docking algorithm — DIFFDOCK[9] that generates and scores binding poses for dopamine receptor active ligands obtained from a ligand-based screening process (Fig 1a).

Despite their potential, these ML methods have yet to be fully validated in experimental settings, particularly in the context of real-world drug discovery. This study seeks to bridge that gap by exploring the effectiveness of these ML approaches in identifying novel agonists for the dopamine D2 receptor, a target of significant therapeutic interest. Dopamine receptors, members of the G-protein coupled receptor (GPCR) family, are crucial regulators of several neurological and physiological processes[11]. Predominantly found in the brain, dopamine receptors are integral to functions such as executive control[12], reward-motivation impulses[13], motor activity[14], and prolactin secretion[15], mediated through four major pathways: mesocortical[16], mesolimbic[17], nigrostriatal[18], and tuberoinfundibular[19]. Dysregulation of these pathways is implicated in a range of disorders, including attention deficit hyperactivity disorder (ADHD)[16], addiction[17], depression[18], schizophrenia[10,18], Parkinson’s disease[18], and hyperprolactinemia[19]. As such, the development of ligands that can selectively activate dopamine receptor subtypes, particularly the D1 and D2 receptors, holds significant therapeutic promise.

Our study focuses on leveraging advanced virtual screening techniques, deploying both discriminative and generative ML models, to discover novel agonists for the D2 receptor. By evaluating these models retrospectively and prospectively, we aim to assess their practical utility and contribute to the ongoing development of more efficient drug discovery pipelines. Through this research, we hope to uncover new chemotypes that could lead to innovative treatments for dopamine dysregulation related disorders.

Figure 1

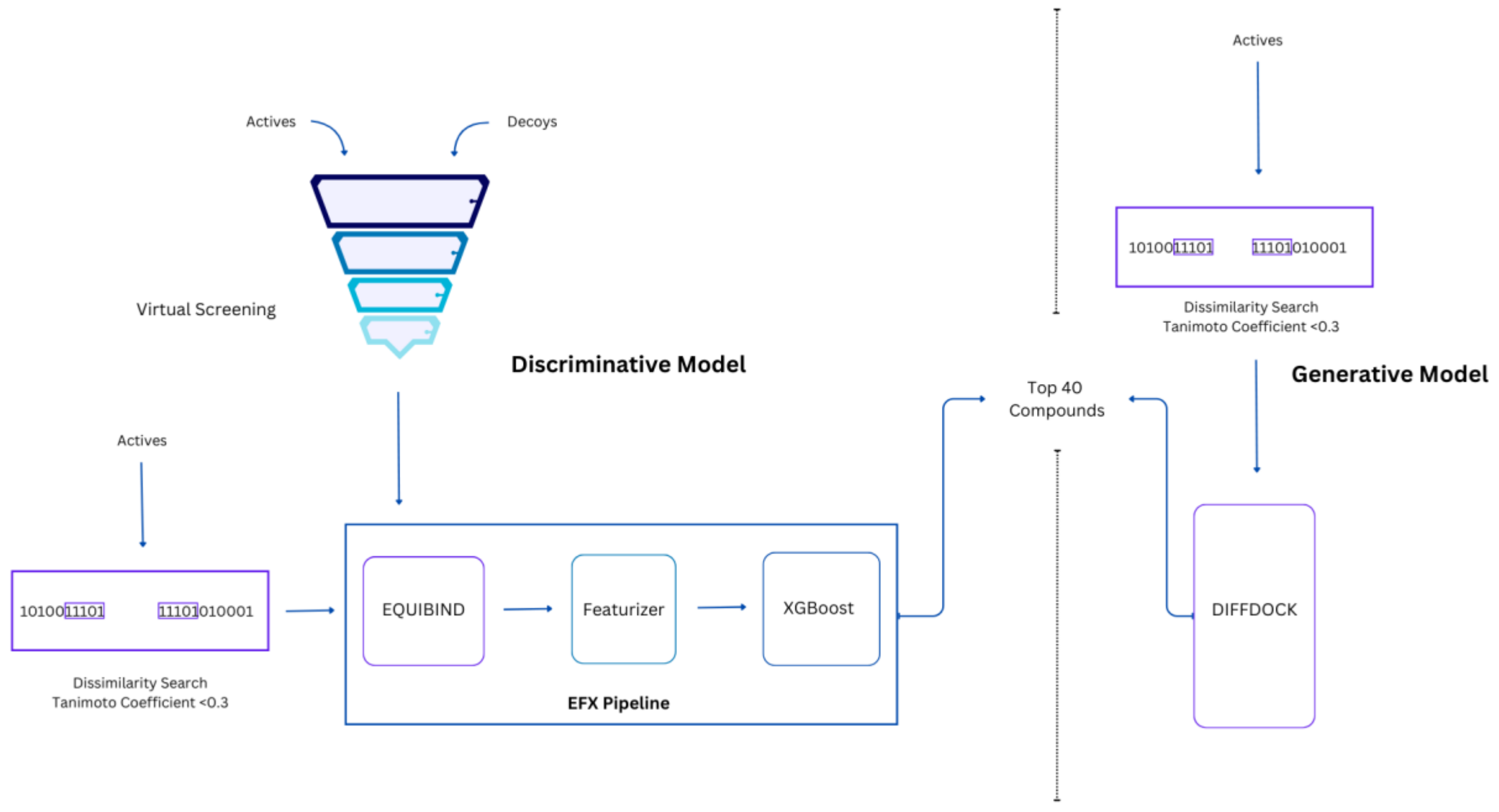

Fig.1 Schematic representation of our workflow. The study's central illustration depicts a Bayesian model approach for virtual screening, integrating discriminative and generative models for binding affinity prediction via EQUIBIND, Featurizer, and XGBoost (EFX pipeline) and DIFFDOCK, respectively.

# Results

## Discriminative Model: EQUIBIND paired with a novel confidence model

Since EQUIBIND was not originally designed for screening, it operates as a regression-based model trained to predict the ground truth coordinates of atoms in a ligand[8]. As such, it lacks an inherent confidence mechanism to directly compare predictions across different ligands. To adapt EQUIBIND for screening purposes, we implemented a complementary confidence model that enables ranking of predictions by their likelihood of being active. Rather than retraining EQUIBIND, our approach focuses on constructing a confidence model capable of discriminating between the predictions generated by EQUIBIND. The discriminative model has three components; EQUIBIND for structure-based docking, an atom-based scoring function, Featurizer used to quantify interactions between atoms in the binding pocket, a classifier model to delineate dopamine receptor actives from decoys (Fig 1a).

The workflow starts by obtaining datasets of active compounds and decoys, curated from BindingDB[20] and the Directory of Useful Decoys-Enhanced (DUD-E)[21], decoys are benchmark datasets which have similar physicochemical properties as actives, however differing in 3D topology[21]. The goal is to remove compounds that do not eliminate compounds that do not conform to the Lipinski's Rule of Five criteria[22,23].
Also, this filtering is crucial to prevent the inclusion of unfeasible or non-drug-like compounds which may introduce outliers, ensuring the actives have similar sample distribution as the decoys. It is important to maintain distribution homogeneity in our training data, to ensure the model is learning the feature patterns and not noise.

**Actives**: The dataset of dopamine D1 and D2 receptor-active compounds includes 3,028 active molecules categorized into 1,426 distinct chemical scaffolds, with 403 identified as agonists and 2,625 as antagonists. The compounds display a wide range of physicochemical properties: molecular weights span from 101.12 to 1721.66 Da (mean 371.44 ± 1.82 Da), lipophilicity (LogP) ranges from -1.74 to 11.95 (mean 3.66 ± 0.22), hydrogen bond donors vary from 0 to 21 (mean 1.21 ±0.02), and hydrogen bond acceptors range from 0 to 19 (mean 4.25 ± 0.03). The molecules have 0 to 55 rotatable bonds (mean 3.98 ± 0.05) and 0 to 9 rings per molecule (mean 3.94 ± 0.02) (fig.2a and Supplementary Fig.1a). Using Lipinski's criteria[23] (fig.2b) focusing on molecular weight and LogP, 1,791 compounds were selected from the dataset. Screening for hydrogen bond characteristics led to the exclusion of 429 active compounds, reducing the number to 1,362, which was then randomly truncated to 1,000 to create a more manageable training dataset for a classical machine learning model (fig.2c, Supplementary Fig.1b ). This refined set was designed to adhere closely to lead-likeness and druggability standards, aiming to enhance model accuracy for distinguishing between active compounds and decoys.

**Decoys**: The DUD-E server was employed to generate a decoy dataset matching 3,028 actives with 50 decoys each, yielding 151,400 decoys with physicochemical properties similar to the actives but different in topological features, intended for a library used in machine learning model development. Physicochemical analysis of the decoys revealed average LogP of 2.57 ± 0.0024, TPSA of 52.80 ± 0.053, molecular weight of 316.23 ± 0.122, and ranges in these properties highlighted their diversity (fig 2d,2e, Supplementary Fig.1c)

**Molecular descriptor and diversity analysis:** To discern the underlying associations among molecular descriptors within the sampled actives and decoys, we employed Principal Component Analysis (PCA)[24] and correlation analysis. Principal Component Analysis (PCA) was conducted on various physicochemical properties of a molecular dataset, including LogP, molecular weight, hydrogen bond acceptors (HBA), hydrogen bond donors (HBD), rotatable bonds, and ring count, to discern their contribution to the dataset's total variance. The PCA, implemented using the Python sklearn library, resulted in seven principal components (PC1 through PC7), with PC1 accounting for 40.65% of the variance, highlighting its dominance in representing the dataset's diversity (Supplementary Fig.1d). Molecular weight and ring count showed significant positive loadings in the PCA, indicating a major influence on the dataset's variance, whereas HBD exhibited a negative loading, suggesting an inverse relationship with the variance (fig .2f)

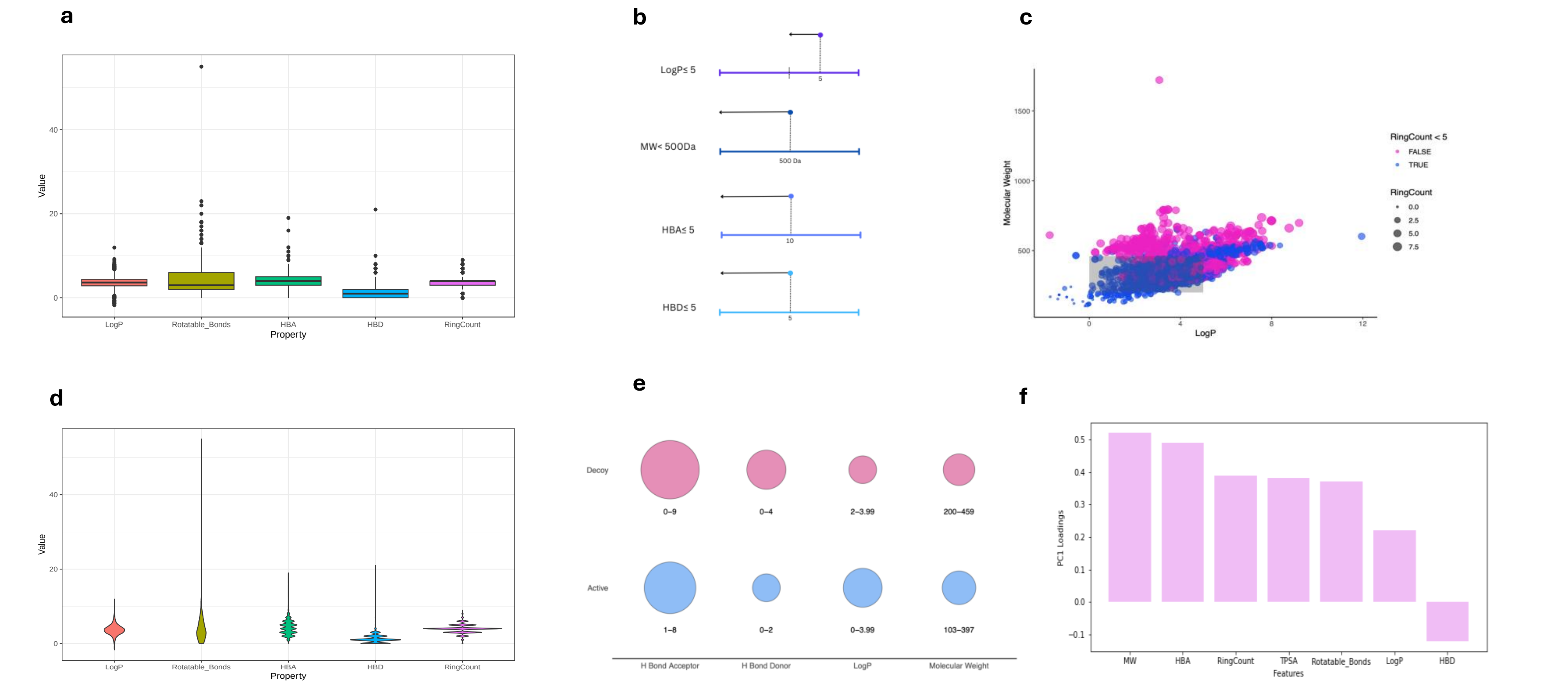

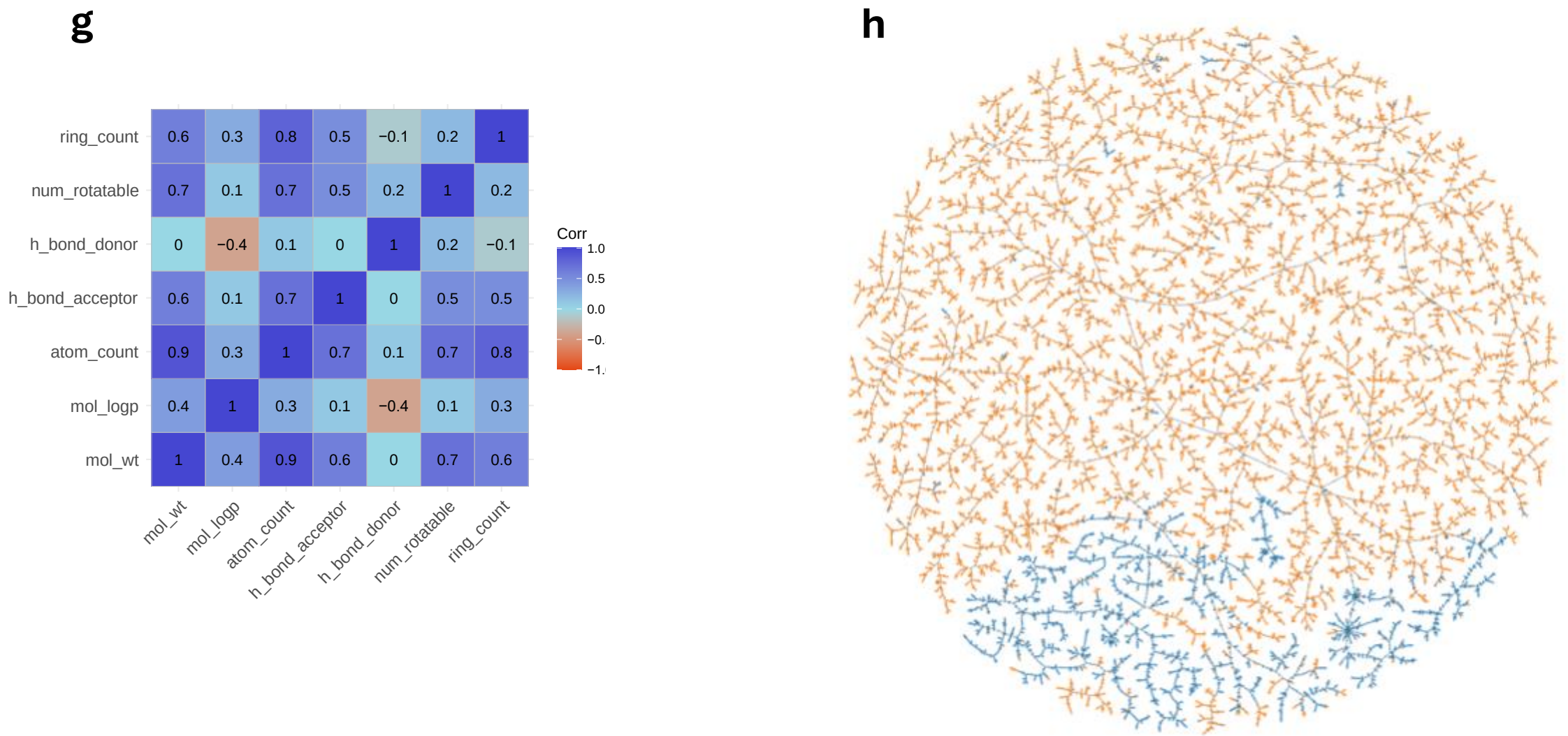

**Figure 2**. **Components and results for discriminative model. a,** A boxplot representing the distribution of actives physicochemical properties; lipophilicity (LogP), number of hydrogen bond donors (HBD), number of hydrogen bond acceptors (HBA), rotatable bond count, ring count for Actives (n= 3028). **b**, Evaluation of compound properties according to Lipinski's Rule of 5. The visual representation includes four metrics: LogP (lipophilicity) ≤ 5, molecular weight (MW) < 500 Da, hydrogen bond acceptors (HBA) ≤ 10, and hydrogen bond donors (HBD) ≤ 5. **c,** Scatter plot showcasing the molecular weight, lipophilicity (LogP), and number of ring structures in the active compounds data. A transparent rectangle highlights the Lipinski range for molecular weight (200-460 Da) and LogP (< 5), emphasizing the compounds that fall within the desirable range for drug-likeness.
**d**, Violin plot representing the distribution of physicochemical properties for decoys (n=151400). **e**, Distribution of molecular descriptors for final datasets of actives and decoys (n=10000). **f**, Loading plot for Principal Component 1 (PC1). This plot showcases the impact of individual molecular descriptors on overall variance. **g,** Correlation matrix displaying the interdependent relationships between molecular descriptors. **h,** Clustered tree-map visualization of actives and decoy scaffold distribution. The map shows the clustering of molecular scaffolds, with active compounds represented in blue and decoy compounds in orange. (Link to interactive tmap: https://temitope-sobodu-dopaminergic-hit-discovery-project.pages.dev/).

The correlation matrix analysis indicates a strong positive correlation between molecular weight and atom count (0.9) and between molecular weight and lipophilicity (0.4), consistent with the idea that larger molecules often exhibit increased hydrophobic characteristics and flexibility due to more rotatable bonds[25]. Conversely, the weak correlation between hydrogen bond donors and acceptors (0) and a negative correlation between lipophilicity and hydrogen bond donors (-0.4) highlight the complex, independent interactions of these properties within the dataset, aiding in feature selection for predictive model development (fig .2g). To evaluate the diversity in a training dataset containing 6398 unique scaffolds (591 actives and 6347 decoys), t-distributed Stochastic Neighbor Embedding (t-SNE)[26] was utilized for its efficacy in reducing dimensionality and visualizing complex data. Using the t-SNE model with parameters set for a perplexity of 30, a maximum of 1000 iterations, and a learning rate of 3, the scaffolds were clustered and visualized using the faerun Python package, revealing a wide-ranging representation within the dataset which is critical for identifying novel ligands in virtual screening (fig.2h,link to interactive tmap: https://temitope-sobodu-dopaminergic-hit-discovery-project.pages.dev/ )

Docking and Scoring: The docking results using EQUIBIND demonstrated that most active compounds align within the 2-5 Å RMSD range (fig.3a), indicating a close approximation to the expected binding pose. While RMSD is useful for evaluating structural alignment, is limited as it does not consider other crucial stereochemical factors like steric clashes, changes in the binding pocket, or electronic interactions, which are essential for accurate binding affinity prediction[27]. To address these limitations, an atom-based scoring function, Featurizer was employed to quantify atomic interactions within a 12 Å radius of the binding site, capturing interactions among various atoms such as Carbon, Oxygen, Nitrogen, Sulfur, and others including Fluorine, Phosphorus, Chlorine, Bromine, and Iodine. This scoring function identified 42 distinct interatomic contacts, excluding hydrogen interactions to reduce noise, enhancing the model's ability to predict binding affinity more reliably. Atomic pair contacts between protein and ligand atoms were quantified, showing carbon-carbon (C-C) interactions as the most frequent for both active and decoy molecules, indicative of prevalent hydrophobic interactions crucial for the stability of protein-ligand complexes (Table 1, Supplementary Fig. 2a,b). The data shows active molecules generally exhibit more extensive atomic interactions than decoys, except for the carbon-sulfur (C-S) pair (Table 1), and these data aid machine learning models in identifying interaction patterns to predict binding affinities with higher precision.

Machine Learning Classification: Machine learning was employed to develop a classification model trained on scored compounds to discern consistent interaction patterns between actives and decoys. Utilizing Logistic Regression, Decision Trees, Random Forest, SVM, and XGBoost. XGBoost outperformed Logistic Regression, SVM, Decision Trees, and Random Forest with an AUROC of 71%, highest accuracy of 95.7%, and an optimal balance of sensitivity (77.1%) and specificity (98.9%), demonstrated by an F1 Score of 0.8, demonstrating superior capability in differentiating active compounds from decoys (Table 2, Supplementary Fig.3a). The confusion matrix for XGBoost showed 165 true positives and 1752 true negatives, with a specificity of 0.981 and 49 false negatives (Supplementary Fig.3b), indicating the model's strong predictive performance yet highlighting areas for sensitivity improvement. The model's early convergence during training and increased complexity indicates a robust capacity for learning intricate patterns, and a potential for overfitting if training is prolonged, which could hinder its generalizability in accurately predicting active compounds and decoys[28,29] (Fig.3c). The model's initial rapid learning and subsequent plateau suggest that further iterations may not substantially enhance its predictive performance for novel molecular entities[29] (Supplementary Fig.3c). The XGBoost model's feature importance revealed that atomic contacts involving Oxygen-Carbon (O.C) and Carbon-Carbon (C.C) are critical for distinguishing active compounds from decoys, whereas sulfur contacts (S.BR, S.F, S.CL) showed less influence (fig. 3d). Interestingly, SHAP[30] (Shapley additive explanations) values reveal features like O.C and C.C significantly impact predictive outcomes, with the frequency of O.C interactions inversely affecting accuracy, while C.C contacts positively correlate with improved performance (fig.3e, Supplementary Fig.3d), enhancing model interpretability through detailed feature contribution insights. This differentiation in feature importance, underscored by SHAP value distributions, enhances the model's interpretability and guides the refinement of molecular descriptors to improve discrimination between actives and decoys. After the machine learning model's training phase was completed, the XGBoost algorithm was selected to complete the EQUIBIND-Featurizer-XGBoost pipeline, which we codified as the EFX pipeline.

## Ligand Based Approach: Screening Enamine's 40 billion compound library via Infinsee

A total of 200 active compounds were selected based on their half-maximal effective concentration (EC50) ranging from 1.86 pM to 0.0104 nM, and half-maximal inhibitory concentration (IC50) spanning 0.21 nM to 0.36 nM, with each compound characterized using Extended Connectivity Fingerprint (ECFP4) within a 4Å radius. These actives were compared against the Enamine RealSpace's repository of over 36 billion compounds using Infinisee software[31], which was calibrated to match each active with 50 ligands having distinct chemotypes and a Tanimoto similarity score below 0.3, aiming to discover ligands with novel scaffolds (Supplementary Fig.3e). After screening, 10,000 compounds were collected and processed through the EFX pipeline, which involved docking, scoring, and classifying compounds as actives or inactives for the dopamine receptor. After LBVS, 10,000 compounds were docked and scored using the EFX pipeline. Compounds stratified by model probability predictions ranging from 0 to 0.995, with those in the Top-1 category (>99th percentile) showing scores from 0.803 to 0.995 (fig. 3f).

## Generative model for binding affinity prediction

DIFFDOCK[9], a diffusion generative model for structure-based docking and ligand pose prediction, treats docking as a generative modeling challenge, employing a diffusion process over the manifold of ligand poses defined by their degrees of freedom (translation, rotation, and torsion). This method maps complex ligand poses to a product space, allowing efficient generation of poses evaluated against target protein structures, and notably surpasses traditional methods like AUTODOCK VINA[32] and GLIDE[33] in accuracy. It maintains superior performance even on computationally folded protein structures, offering fast inference times (average 40 seconds per inference) and providing confidence estimates with high selective accuracy.

Initially, a benchmark consisting of 100 known actives and 100 decoys was established through docking with DIFFDOCK to serve as a reference for model evaluations and to assess its generalizability. The results showed that D1 receptor actives had a high normalized confidence average of 0.94 ± 0.06, while decoys averaged lower at 0.45 ± 0.13 (Supplementary Fig.3f), indicating the model effectively differentiated actives from decoys based on their binding probabilities and aligned closely with empirical data.

DIFFDOCK was used to dock 10,000 compounds, sourced from ligand-based screening, onto the co-crystal structures of the Dopamine D1 and D2 receptors, offering a built-in confidence metric for evaluation. For each compound, DIFFDOCK produced 40 potential receptor-ligand binding poses, assigning scores ranging from 0 to 39, where a score of 0 represents the most confident prediction of the receptor-ligand pose. We obtained results for the Top-1 receptor-ligand poses, employing a confidence metric that approaches zero for the highest potential binding affinity (fig. 3g). This metric was normalized. The min-max normalization converted the confidence metric to a 0-100% scale, with 100% representing the maximal binding affinity potential. These normalized scores are comparable to the probability scores from the EFX pipeline, allowing for a direct comparison of the Top-1 potential hits from both generative and discriminative models. Within the DIFFDOCK confidence metric, a total of 258 compounds achieved a perfect score of 100%, and 3,062 compounds were placed within the 95th to 99th percentile range. In contrast, the EFX model's probability scores were more conservative, with only 11 compounds receiving probability scores above 90%. the top 20 compounds were shortlisted from the Top-1 predictions of both models for post-modeling screening (Fig. 3g).

## Post-Modeling Screening

The Top 20 predictions from the discriminative and generative models were combined, with DIFFDOCK consistently achieving a confidence level of 100, while scores from the EFX pipeline ranged from 88.57 to 99.95. Most of the 40 compounds evaluated conformed to Lipinski's rules of five[23], with characteristics including a molecular weight range from 128.12 to 582.21, a LogP range from -2.34 to 5.07 (average 2.05), a TPSA from 13.67 to 109.07, and maximum hydrogen bond donors and acceptors capped at 3 and 5, respectively (fig. 3h). These properties align with recognized pharmacokinetic benchmarks and, assuming these molecules activate dopamine receptors, enhance their potential for pharmacodynamic efficacy. Compounds 6, 15, 38, and 40 were noted for their LogP values being less than 0, indicating high polarity, which could potentially compromise their ability to penetrate the brain. Compound 9, with a molecular weight over 500 Da, were excluded due to their potential inability to penetrate the brain, crucial for targeting dopamine receptors behind the blood-brain barrier (BBB). The BBB restricts the entry highly polar or large ligands[34], emphasizing the importance of adhering to Lipinski's rules to enhance the likelihood of central bioavailability.

**Table1**. Top 10 atomic pair contacts by counts ($Atom_{protein} – Atom_{ligand}$)

| Atomic Pair Contact | Actives Average | Decoys Average | Combined Maximum | Combined Average |
|---|---|---|---|---|
| C-C | 4441 | 3840 | 17001 | 3904 |
| O-C | 1125 | 969 | 4098 | 985 |
| N-C | 972 | 838 | 3575 | 852 |
| C-N | 643 | 625 | 3217 | 627 |
| C-O | 465 | 372 | 3438 | 382 |
| O-N | 164 | 159 | 747 | 160 |
| N-N | 142 | 137 | 666 | 138 |
| C-S | 50 | 108 | 1033 | 101 |
| O-O | 1186 | 94 | 830 | 96 |
| N-O | 102 | 81 | 731 | 83 |

Compounds 2, 10, 17, 27, 12, 13, 16, 19, 23, and 35 were excluded from the cohort due to their inclusion of chemical groups associated with significant toxicity risks, including halogenated aromatics known for endocrine disruption and carcinogenicity, quinoline and hydroxylamine groups capable of forming reactive metabolites, and polycyclic and macrocyclic structures with classic toxicophore properties[35,36]. Additionally, azides in Compound 35 were specifically removed due to their high reactivity potentially leading to explosive reactions and their prevalent use in bioconjugation chemistry involving click chemistry reactions[37]. We employed the Aggregator Advisor tool (available at https://advisor.bkslab.org/)[38] to detect colloidal aggregators in our compounds, with Compounds 11, 32, and 33 being flagged as potential aggregators. While this screening tool does not completely guarantee the elimination of frequent hitters or false positives, it greatly assists in identifying compounds that share characteristics or chemical profiles with known assay disruptors[39,40]. To assess the novelty of the top 40 predictions, molecular scaffolds were generated using the Bemis-Murcko function in RDKit and compared against 1,457 unique scaffolds from 3,028 known dopamine receptor actives, revealing that the top predictions introduced 40 distinct scaffolds. Tanimoto coefficients were calculated to measure similarity between these new scaffolds and existing ones, with the highest similarity coefficient found to be 0.32 (fig. 3i), indicating a significant structural distinction from known compounds and highlighting the innovative potential of the predicted compounds for targeting dopamine receptors. After passing the screening, the 21 selected compounds were sent to Enamine, the supplier of the virtual library RealSpace where the molecules were sourced through ligand-based screening, for synthesis. Enamine confirmed unsuccessful synthesis attempts for compounds 4, 5, 14, 18, 30, and 37, whereas compounds 1, 7, 8, 20, 21, 22, 24, 25, 26, 28, 29, 31, 34, 36, and 39 were successfully synthesized with each achieving a purity level exceeding 95%.

**Table 2. Summary of model evaluation metrics**. The XGBoost model demonstrates superior performance with the highest accuracy of 95.7% and a balance between sensitivity and specificity (77.1% and 98.9%, respectively), the model's F1 Score of 0.8.

| Model | TP | FP | FN | TN | Accuracy | Sensitivity | Specificity | Precision | RMSE | F1 |
|---|---|---|---|---|---|---|---|---|---|---|
| Logistic Regression | 1 | 10 | 213 | 1776 | 0.885 | 0.005 | 0.994 | 0.091 | 4.768 | 0.009 |
| SVM | 6 | 2 | 208 | 1784 | 0.899 | 0.257 | 0.977 | 0.750 | 4.652 | 0.018 |
| Decision Trees | 32 | 42 | 182 | 1744 | 0.888 | 0.151 | 0.977 | 0.432 | 4.176 | 0.222 |
| Random Forest | 28 | 54 | 186 | 1732 | 0.88 | 0.13 | 0.961 | 0.341 | 4.33 | 0.189 |
| XGBoost | 165 | 34 | 49 | 1752 | 0.957 | 0.771 | 0.989 | 0.829 | 1.333 | 0.8 |

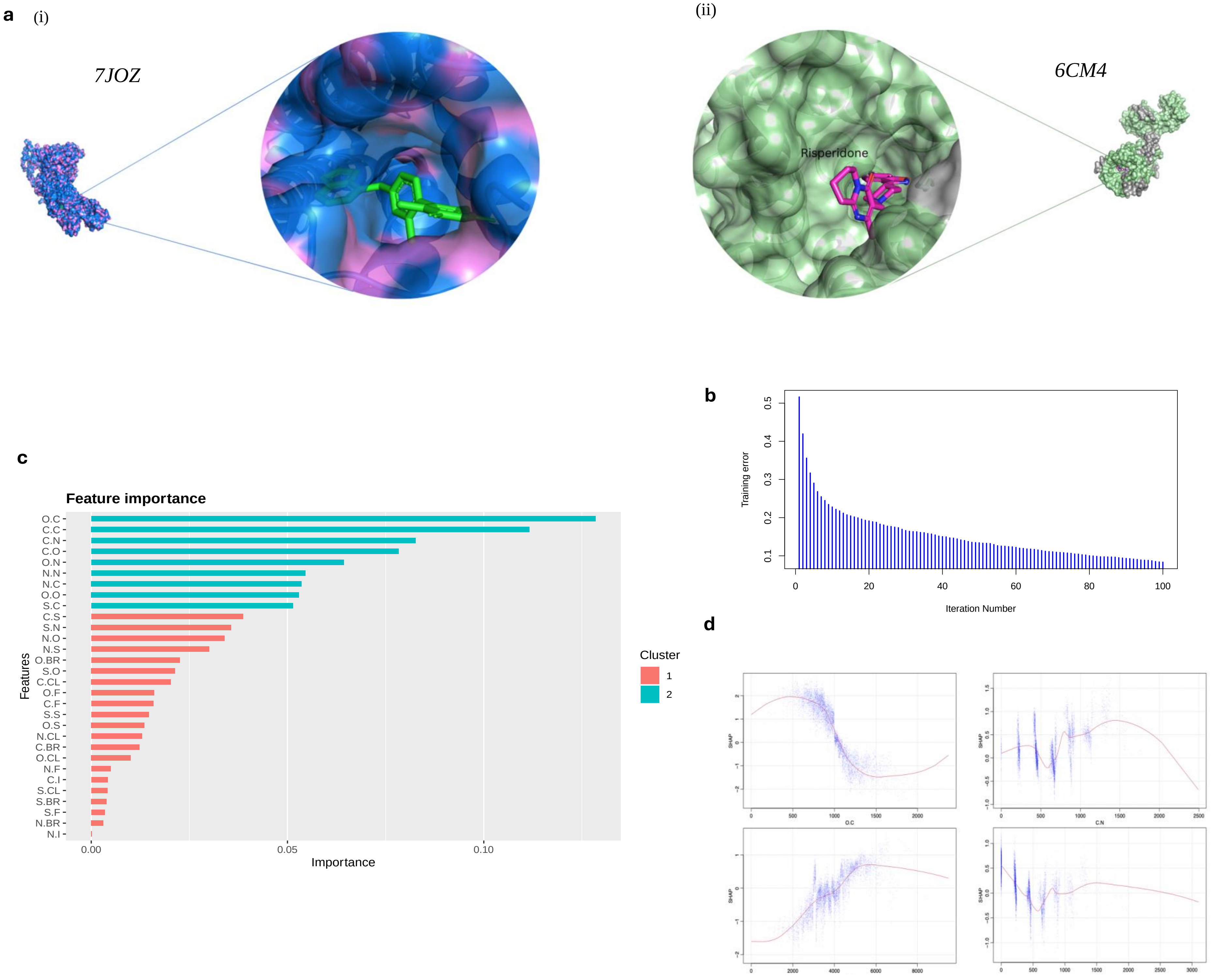

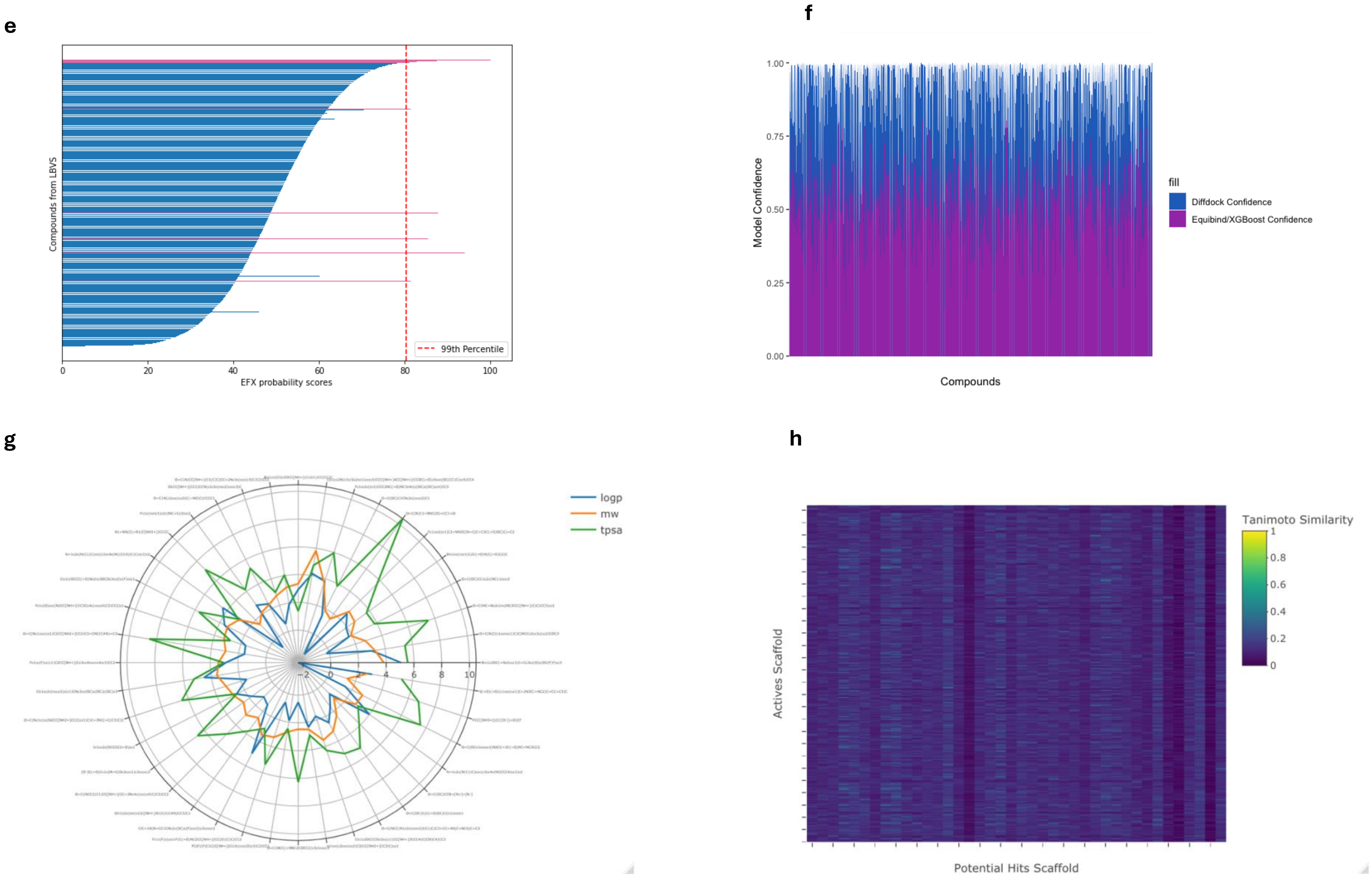

**Figure 3.** **Components and results for generative model. a,** Co-crystal structures of the dopamine D1 (PDB: 7JOZ) in complex with a G protein and a non-catechol agonist, and dopamine D2 receptor (PDB: 6CM4) in complex with risperidone. The image shows an overall view of the receptor structure, with a zoomed-in section highlighting the binding site of the ligand. **b,** XGBoost model iterations (rounds = 100) calibrated to the training error (mean: 0.153). **c,** Feature importance plot: Weight of atomic contacts on model prediction. **d,** SHapley Additive exPlanations (SHAP) for Oxygen-Carbon (O-C), Carbon-Nitrogen (C-N), Carbon-Carbon (C-C), and Carbon-Oxygen (C-O) atomic interactions. **e,** Distribution of predicted probability scores from the XGBoost model for potential hits classification. The plot shows the distribution of predicted probability scores, with each horizontal line representing a compound. The top 1% of compounds, classified as potential hits, are highlighted in red. The red dashed line indicates the 99th percentile threshold, separating the top-scoring compounds from the rest. **f,** Min-Max normalization of the DIFFDOCK confidence metric, corresponding the with the EFX pipeline probability score. **g,** A radar chart illustrating the normalized distribution of the Molecular Weight, LogP and TPSA of the top 40 predicted compounds(link to interactive chart: https://temitope-sobodu-dopaminergic-hit-discovery-project.pages.dev/selected40). **h,** Heat map comparing the Tanimoto coefficients of active scaffolds (n = 1457) and predicted compounds scaffolds (n = 40). link to interactive chart https://temitope-sobodu-dopaminergic-hit-discovery-project.pages.dev/tani_simi_heat).

## Validation

**Binding Assays**: Using genetically encoded biosensor constructs—dLight[40,41]—in a live cellular imaging setup, the dLight sensor was introduced into HEK293A cells via adenoviral plasmids and imaged using a Nikon TiE Eclipse confocal microscope to allow high-resolution visualization of dopaminergic activation while preserving spatiotemporal dynamics. Initially, compounds are tested against dLight1.1, specific to the dopamine D1 receptor; those failing to activate dLight1.1 are then screened against dLight1.5, specific to the D2 receptor, to identify potential D1 activators, selective D1 ligands, and D2 receptor ligands. At the conclusion of this screening, compounds are classified as selective D1 ligands, selective D2 ligands, or non-selective dopaminergic ligands, depending on their activation of dLight1.1, dLight1.5, or both (fig. 4a). The assay calibration utilized a negative control (GFP and vehicle) and dLight1.1 constructs exposed to varying dopamine concentrations, resulting in fluorescence intensity changes that peaked at a 67% increase at 10 μM dopamine, indicating receptor activation with a calculated Kd of 0.077±0.014 μM (Supplementary Fig.4a). Fenoldopam, a selective D1 agonist, mirrored dopamine's effects at 10 μM with a similar fluorescence change and a Kd of 0.108±0.04 μM, while the D1 antagonist SCH23390 effectively abolished these responses, confirming competitive inhibition with a Ki of 0.093 μM (Supplementary Fig.4b, c). During in vitro screening of the shortlisted 21 compounds, only Compound 8 activated dLight1.1 at 10 μM (Kd = 9.91±0.19 μM), demonstrating lower potency compared to dopamine and fenoldopam and suggesting that higher-affinity antagonists could displace it more readily at the D1 receptor (Supplementary Fig.4d). Further, the top 21 compounds were tested for D2 receptor activity using the dLight1.5 biosensor in HEK293A cells. Bromocriptine exhibited a peak fluorescence intensity of 211.23 ± 7.28, higher than dopamine's peak of 167.82 ± 13.2 at 1 μM, with both peaks effectively abolished by the D2 antagonist haloperidol at 20 μM, indicating a competitive antagonistic interaction, while the D1 antagonist SCH23390 at 1 μM did not affect these responses (Supplementary Fig.4e,f), confirming the specificity of dLight1.5 for D2 receptor ligands. Compounds 1 and 34 activated the dLight1.5 sensor (Kd = 0.104± 0.144 $\mu$M and Kd = 0.153± 0.039 $\mu$M respectively), demonstrating fluorescence intensities at sub-micromolar concentrations comparable to dopamine and were effectively suppressed by haloperidol, suggesting competitive interactions with the D2 receptor (fig. 4c,d). These compounds showed high potency and strong binding affinity to the D2 dopamine receptor, indicated by robust activations with fluorescence fold changes of 77.36% and 71.92% respectively, while minimal activation of the RdLight1 sensor for D1 receptors at 4.3% underscores their selectivity for D2 receptors (fig. 4e,f)

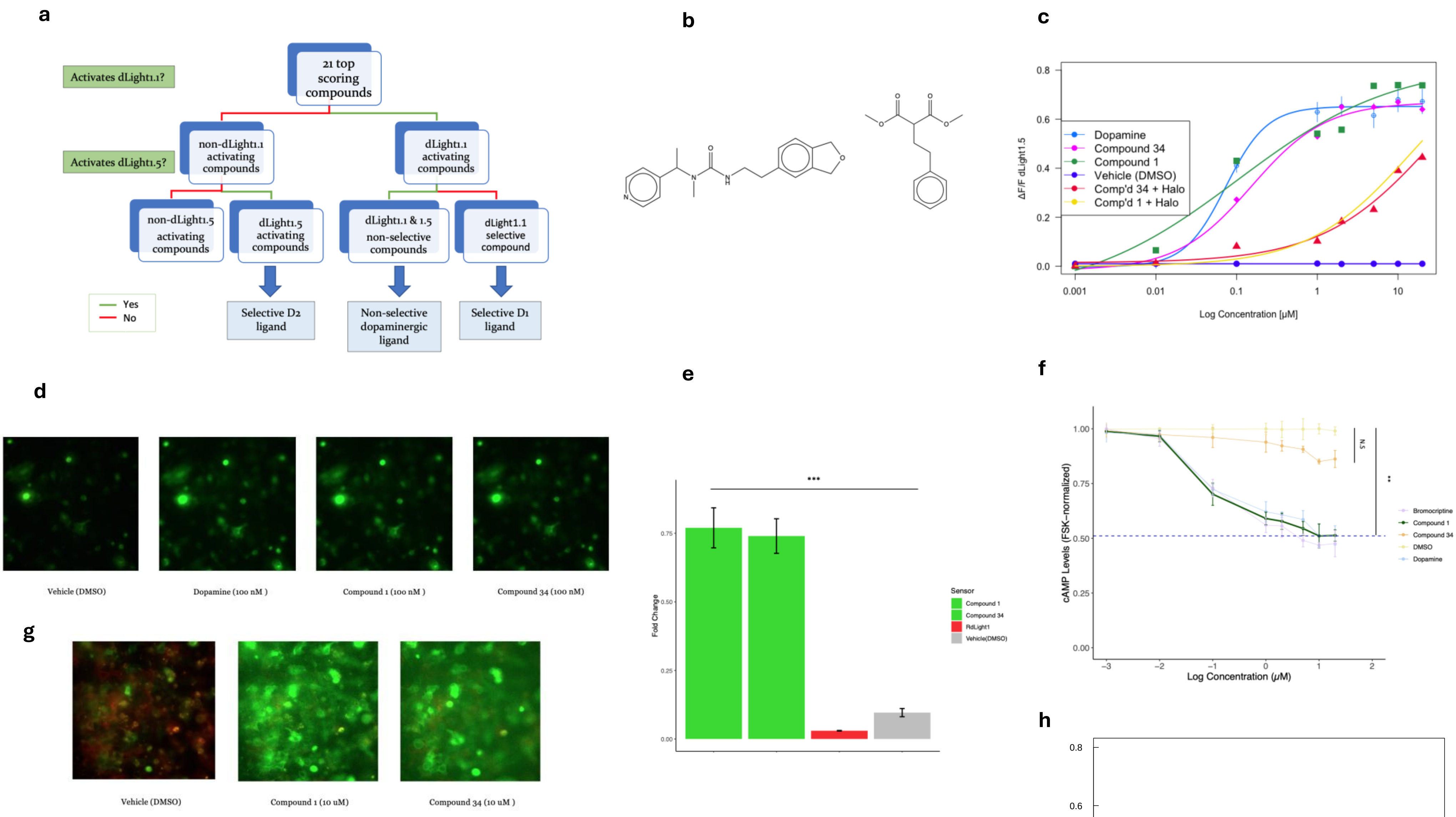

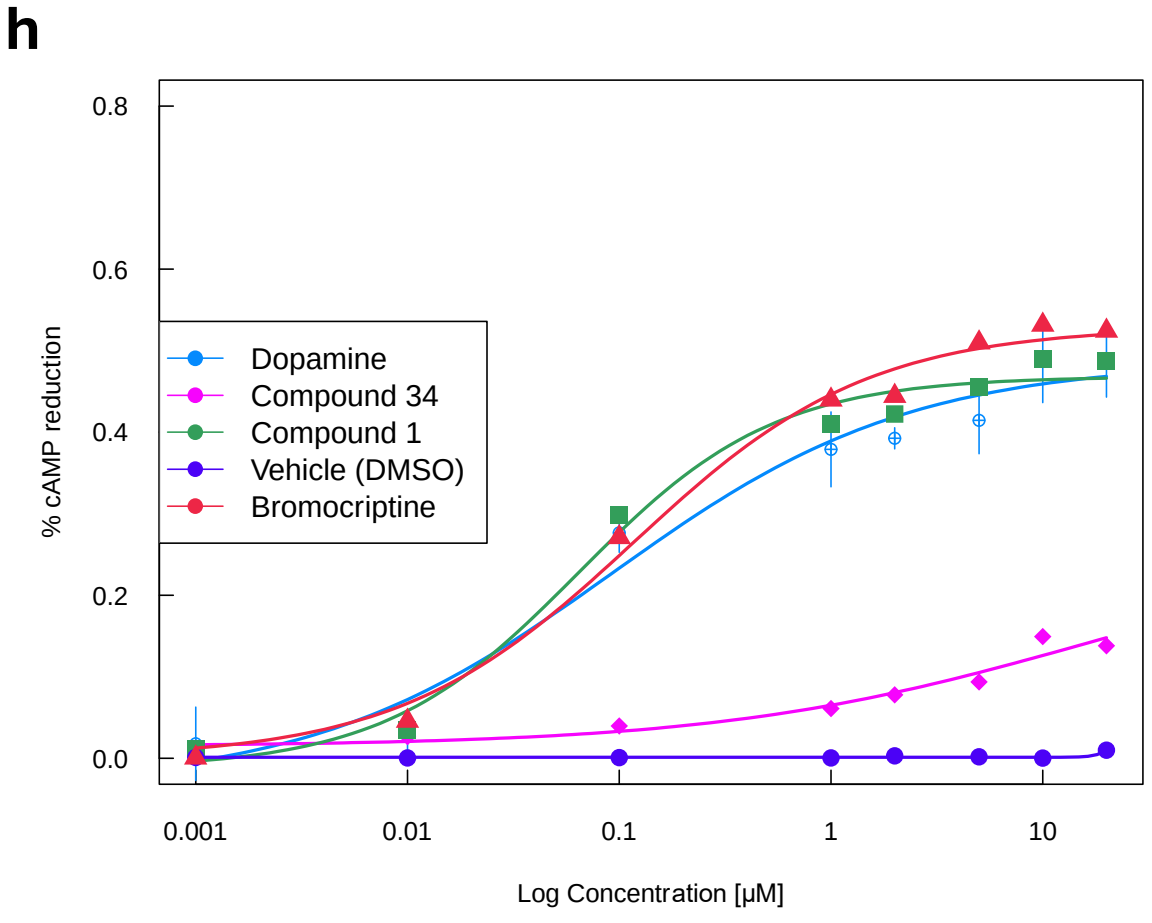

**Figure 4. Validation with binding and functional assays. a,** In Vitro validation flow chart. This chart outlines the validation process for the top 21 scoring compounds based on their activation of dLight1.1 and dLight1.5 (dopamine D1 and D2 cpGFP modules respectively). **b,** Chemical structures for compound 1 (left) and compound 34 (right). **c,** Titration of graded concentrations Compound 1 (Kd = 0.104± 0.144 $\mu$M) and Compound 34 (Kd = 0.153± 0.039 $\mu$M) on HEK 293A cells expressing dLight1.5 with pre-treated Haloperidol (20 $\mu$M) (mean ± SEM). Fitted with Hill equation. **d,** Live cellular imaging shows the activation of dLight1.5 in HEK293A cells by sub-micromolar concentrations of dopamine (100 nM), compound 1 and 34 (100 nM) [Link to imaging video](). **e,** Live cellular imaging shows the activation of dLight1.5 (green) and RdLight1 (red) in HEK293A cells by compound 1 and 34 (5 $\mu$M). **f,** Quantification of fold change in dLight1.5 and RdLight1 fluorescence in the presence of compounds 1 and 34 (n = 5 sessions). Data shown are means ± SEM. *P < 0.05, **P < 0.01, ***P < 0.001; one-way ANOVA analysis was done followed with a Dunnett's post hoc test. **g,** cAMP response curve to a titration of compounds 1, 34, dopamine, and bromocriptine in SH-SY5Y cells expressing wild-type DRD2 receptor. (n=3, **p<0.01, ***p<0.001, one-way ANOVA, Dunnett's post-hoc test) [Link to imaging video](). **h,** Effective Concentration plot of graded concentrations of Compound 1 (EC50 = 0.064± 0.02 $\mu$M) and Compound 34 (EC50 = 1.54 ± 8.48 $\mu$M) on SH-SY5Y cells expressing DRD2 (mean ± SEM). Fitted with Hill equation.

**Functional Assays**: Upon ligand binding, the D2 dopamine receptor couples with the Gi protein subunit, leading to adenylyl cyclase inhibition and a decrease in cyclic Adenosine Monophosphate (cAMP) production, indicating receptor activation essential for intracellular signaling. Using the GloSensor plasmid cAMP reporter with D-luciferin in SH-SY5Y cells expressing the D2 receptor, compounds 1 and 34 were tested alongside forskolin, an adenylyl cyclase activator; compound 1 significantly suppressed cAMP levels by 52.87%, similar to dopamine and bromocriptine, while compound 34 showed only a modest decrease. These results confirm compound 1's efficacy as a D2 agonist with an EC50 of 64.32 nM and suggest that compound 34, despite high receptor affinity, lacks full agonist properties with an EC50 of 1.54 µM (fig.4g,h).

## Discussion

The complementary strengths of LBVS and SBVS have been instrumental in expanding the horizons of identifying novel therapeutic candidates[42,43]. In this study, the synergistic integration of LBVS and SBVS techniques has enhanced discovery outcomes. This combination strategy capitalizes on the broad search space offered by LBVS and the precision of SBVS in affinity prediction[43,44]. The dual approach helps in circumventing the limitations inherent in each method when used in isolation, thereby increasing the robustness and success rate of the screening process[42,43,44].
Furthermore, we explored a Bayesian paradigm in modeling our structure-based methods by constructing a discriminative and generative machine learning pipeline. Our discriminative model was based on EQUIBIND[8], a geometric graph neural network for docking ligands and predicting binding poses that employs a regression-based framework. Although EQUIBIND, in its original form, has limited capabilities (5.5% accuracy for Top-1, RMSD <2Å predictions) due to its inability to model non-static components of the ligand structure, such as side chains[9], we enhanced performance by co-opting the model as a feature extraction engine. These features were then scored using Featurizer[10], an atom-type-based scoring function. Featurizer is a knowledge-based protein-ligand scoring function that aggregates paired atomic contacts within a 12Å radius of the binding epicenter[10]. Leveraging the atomic pair counts data, we classified actives and decoys using classical machine learning methods. Our XGBoost model demonstrated superior performance with an F1 score of 0.8. The discriminative model (EFX pipeline) predicted compound 8, which binds to the dopamine D1 receptor with micromolar affinity (>10 µM). For the generative model, we deployed DIFFDOCK, which used a diffusion-based approach to generate conformers and poses in the manifold, differing from earlier regression-based binding pose estimations with limited conformers[9]. DIFFDOCK learns by denoising and reconstructing signals to produce multiple conformations, scoring these conformations using a confidence model trained on cross-entropy loss[9]. The LBVS-DIFFDOCK pipeline produced compounds 1 and 34, both with novel scaffolds and nanomolar affinity for the dopamine D2 receptor, screened using dLight sensors. Compound 1 demonstrated a reduction in cAMP accumulation similar to dopamine and bromocriptine. Compound 34 did not replicate this response, suggesting it might not engage the canonical Gi-Adenylyl Cyclase inhibition pathway and may promote biased dopaminergic downstream signal transduction. A limitation we encountered is that our structure-based models were non-state-specific for the protein-ligand conformation. Besides inherent model accuracy, this contributes to the variability in results for both the discriminative and generative models, despite working on the same dataset. Our methods explored the Bayesian modeling dichotomy without assessing superiority, aiming to shift the focus to outcomes and demonstrate that different modeling pathways can lead to variable, non-stochastic results. Ultimately, we aimed to demonstrate the promise of virtual screening and machine learning in the discovery of compounds with both in silico and in vitro activity.

## Conclusion

This study led to the discovery of a novel dopamine receptor agonist (Compound 1) with a non-catechol scaffold, achieved through discriminative and generative machine learning methods that effectively navigated vast chemical spaces. The integration of ligand-based and structure-based virtual screening facilitated precise predictions and selections based on molecular interactions, demonstrating the synergy of machine learning and virtual screening in drug discovery, and emphasizing the efficiency and specificity brought by these methodologies in identifying therapeutic candidates. We anticipate these combination strategies to become more prevalent in the search for therapeutic candidates for diverse disease burdens.

## Methods

**Dataset provenance**: Datasets of dopamine actives and decoys were sourced from the Binding Database (bindingDB.org)[20] and the DUD-E server (dude.docking.org)[21], respectively. I downloaded a dataset of 3,028 Dopamine D1 actives from BindingDB using EC50 and IC50 as selection parameters, listed ligands with EC50 < 20uM for agonists, IC50 <92.8uM for antagonist. Decoys were generated by inputting Dopamine actives in SMILES format into the DUD-E server, which returned 50 decoys per active (random hit rate: 2%), totaling 151,400 unique molecules, each assigned a distinct decoy ID. To ensure decoy uniqueness, I generated International Chemical Identifier Keys (InChIKeys). Consistent with the original authors of the DUD-E dataset, a compound for a predefined protein target is considered a decoy when its pKi for the target is less than 6. Subsequently, RDKit, an open-source cheminformatics toolkit, was employed to extrapolate the physicochemical properties from the provided SMILES format for actives and decoys. Utilizing computational algorithms, RDKit enabled the calculation of various properties such as Molecular Weight (MW), Log Partition Coefficient (LogP), Total Polar Surface Area (TPSA), alongside the count of Hydrogen Bond Donors, Hydrogen Bond Acceptors, Rotatable Bonds, and Ring structures. Adhered to Lipinski's rule[22,23] of five to filter out non-druggable compounds, retaining only those meeting the established criteria: a molecular weight < 500 Daltons, LogP < 5, fewer than 5 hydrogen bond donors, less than 10 hydrogen bond acceptors, and no more than 10 rotatable bonds. Following filtration, Microsoft Excel was utilized for randomized trimming of the data, aligning the decoy dataset with the actives at a ratio of 10:1, setting our enrichment factor at 10%. The final dataset, prepared for docking, comprises 1,000 actives and 9,000 decoys.

**Exploratory Data Analysis**: In the exploratory data analysis, the aim was to discern patterns and correlations within the molecular descriptors properties of active compounds and decoys. Using t-Distributed Stochastic Neighbor Embedding (t-SNE)[26], the data's dimensionality was reduced. Morgan fingerprints, a way of representing the structure of a molecule in a compact binary form. These fingerprints work by analyzing the molecule's atoms and their connectivity, considering each atom and its neighborhood up to a defined radius, and encoding this information into a fixed-length array of bits (0s and 1s)[45]. The Morgan fingerprints for each molecule was computed using RDKit, with a radius of 2 Å and bit vector size of 2048, capturing the essence of chemical structures for comparison. The multidimensional data was visualized using Python's Faerun package, resulting in an interactive clustered heatmap.

**Docking and Machine Learning**: For docking with EQUIBIND we converted compounds in SMILES TO SDF, OpenBabel[46] was employed. Concatenating the dopamine receptor crystal structures (PDB: 7JOZ for dopamine D1[47], and 6CM4 for dopamine D2[48]) with 10,000 compounds producing protein-ligand complexes tucked in individual folders. Initiated a batch docking process, using a local computing setup with a 2.2 GHz Core i7 processor and 16 GB (Random Access Memory) RAM, resulting in an average run-time of 24 seconds per inference. The output, organized in dedicated directories for each protein-ligand complex, facilitates the subsequent analysis of binding interactions. EQUIBIND evaluates protein-ligand binding poses via three benchmarks: Ligand RMSD, Centroid distance and the Kabsch RMSD[8]. All metrics are outputted in the result folder. Featurizer extracts interatomic contacts within a 12 Å threshold between ligand and protein in the binding pocket, creating a structured dataset detailing the frequency of elemental atomic contacts for subsequent model training and docking pose evaluation[10].

We considered the five common elemental atom types for protein P and nine for the ligand L.

These are:

- For the protein atom types: *P(j) ∈{C, N, O, P, S}*

- For the ligand atom types: *L(i) ∈{C, N, O, F, P, S, Cl, Br, I}*

The occurrence count for a specific protein atom type (*j*) interacting with a specific ligand atom type (*i*) is defined as:

$$[x_{Z(P(j)),Z(L(i))} = \sum_{k=1}^{K_j} \sum_{l=1}^{L_i} \Theta(d_{\text{cutoff}} - d_{kl})]$$

Where:

$d_{kl}$ is the Euclidean distance between the *k-th* protein atom of type
*j* and the *l-th* ligand atom of type *i*.
$K_j$ is the total number of protein atoms of type *j*.
$L_i$ is the total number of ligand atoms of type *i*.
$Z$ is a function that returns the atomic number of the element, and this is used to rename the feature with a mnemonic denomination.
$\Theta$ is the Heaviside step function that counts contacts within a cutoff distance
$d_{cutoff}$ = 12 Å.

To accurately distinguishing between active compounds and decoys within a dataset comprising 10,000 entries (1,000 actives and 9,000 decoys). Classification machine learning models were employed, including Logistic Regression, Decision Trees, Random Forest (RF), Support Vector Machines (SVM), and Extreme Gradient Boosting (XGBoost). The dataset was divided into training and test sets in an 80:20 ratio. To avoid bias towards either actives or decoys in the training and test sets, Leave-One-Out Cross-Validation (LOOCV)[49] was utilized. The dataset consisted of atomic contact pairs as independent variables, represented by elemental symbols from the periodic table (C.C, N.C, O.C, S.C, C.N, N.N, O.N, S.N, C.O, N.O, O.O, S.O, C.F, N.F, O.F, S.F, C.P, N.P, O.P, S.P, C.S, N.S, O.S, S.S, C.CL, N.CL, O.CL, S.CL, C.BR, N.BR, O.BR, S.BR, C.I, N.I, O.I, S.I). The dependent variable was the class of the compound as either active or decoy. Evaluation metrics were Fischer Score (F1) and Matthews Correlation Coefficient. Classification models were constructed for Decision Trees, Random Forest, Support Vector Machines, and Extreme Gradient Boosting. The Decision Tree model employed the rpart function in R, with post-training parameter tuning through tree pruning and a complexity cost set to 0.12. The Random Forest model used 2001 trees and attempted four variables at each split, facilitating Out of Bag error rate estimation post-training. XGBoost was configured with binary logistic objective, utilizing 3 threads, over 100 rounds, with a learning rate of 0.3. The SVM model, with a radial basis function kernel, was parameterized with a cost of 1 and 2001 support vectors. The model with the best evaluation metrics was selected for predicting actives from ligand-based screened compounds.
DIFFDOCK docked 10,000 molecules derived from ligand-based similarity assessments. DIFFDOCK enables the docking of SMILES format compounds onto a protein structure[9], in this case, PDB: 7JOZ for dopamine D1[47] and 6MC4 for dopamine D2[48]. It generated multiple poses for each protein-ligand complex and employs a confidence metric alongside a scoring function to rank the top 40 binding conformations, with a preference for those with RMSD values under 2 Å. DIFFDOCK was ran on an NVIDIA Quadro RTX GPU. Each simulation averaged 82 seconds, with batches of 500 compounds processed sequentially. The results, prioritized from the most to least active, were organized into designated directories.

**Post-modeling Screening**: In the filtration phase, the top 40 scoring ligands were screened to exclude Pan-Assay Interference Compounds (PAINS)[50,51] and frequent-hitters[52]. Colloidal aggregators were identified using the Aggregator Advisor (https://advisor.bkslab.org/)[53], and spectroscopy interference compounds using the Pubchem Bioassay Library (https://www.ncbi.nlm.nih.gov/pcassay)[54]. Potentially toxic functional groups were visually inspected, resulting in 21 compounds free of chemical liabilities, which were advanced to the cellular validation stage for further biological activity assessment. Scaffold novelty for the 21 compounds was determined using the Sci-Finder platform (https://www.cas.org/solutions/cas-scifinder-discovery-platform/cas-scifinder).

## In vitro Validation

**Plasmids**: Constructs pAAV-CAG-dlight1.1, pCMV-dlight1.5, and pAAV-CAG-Rdlight1 (Addgene) containing circular permuted Green Fluorescent Protein (cpGFP) were used for ligand binding assays by fusing with Dopamine D1 and D2 receptors. Negative controls included pAAV-AMPK-DIO-mCherry and pCMV-GFP plasmids, and pCMV-Frankenbody-FLAG-mRuby2 was used to trace membrane expression and internalization of the constructs. Functional assays were done using DRD1 and DRD2 circular DNA plasmids (Bio-techne), monitoring cAMP levels with the pGLOSensor-20F luciferase construct (Promega), with forskolin treatment for D1 and D2 receptor assays. Plasmid integrity was confirmed via southern blotting after agarose gel electrophoresis.

**Bacterial transformation**: The QIAGEN Plasmid Maxi Kit protocol is initiated by selecting a single colony from a newly streaked plate with Luria Broth (LB) media, and a small starter culture by inoculating it into 2–5 mL of LB medium supplemented with Ampillicin (100 ug/mL) for pAAV-CAG-dlight1.1, pAAV-CAG-Rdlight1, pCMV-GFP, Kanamycin (50 ug/mL) for pCMV-dlight1.5, pAAV-AMPK-DIO-mCherry, pCMV-Frankenbody-FLAG-mRuby2. Post incubation, the bacterial cells were harvested by centrifugation and proceeded with the resuspension using Buffer P1, Buffer P2 for lysis, and Buffer P3 for neutralization. Then the application of the clear lysate to a QIAGEN-tip to bind the DNA, which was subsequently washed and eluted. The extracted DNA was precipitated using isopropanol, washed with ethanol, and air-dried before being resuspended.

**Cell Culture and transfection**: Human Embryonic Kidney Cells Subclone A (HEK293A) was used for *in vitro* validation of selected ligands from the virtual screening stage. The Human Embryonic Kidney 293 (HEK293A) cells, available through ATCC (catalog number: PTA-2511). The maintenance protocol involves thawing the cells rapidly at 37 °C and transferring them to growth media post-decontamination. Cells are cultured in a 37 °C, 5% CO2 incubator, with passaging required twice weekly. For confluency, a T75 flask (Thermofisher) typically reaches 1 × 10^7 cells. During passaging, cells are washed with Phosphate Buffer Solution (Gibco), treated with trypsin (Sigma-Aldrich), and then counted using a hemacytometer. Resuspended cells are seeded at appropriate densities for continued culture. The freezing process involves growing cells to 50% confluency, replacing growth medium, and using a cryoprotective freezing medium before storage at –80 °C, followed by transfer to liquid nitrogen for long-term storage. For the transfection of cDNA into HEK 293A cells, a 24-well glass bottom plate was utilized to culture the cells, which were then dissociated with 2.5% Trypsin after a 24-hour incubation. Optimal cell density of 50,000 to 125,000 cells per well in 500 μL of complete medium to reach 50-80% confluence was ensured on the day of transfection. Afterwards the preparation of the transfection mixture was done by diluting Lipofectamine 3000 reagent and plasmid DNA in separate aliquots of Opti-MEM Reduced Serum Media, incorporating the P3000 reagent according to the DNA quantity. Post mixing, the solutions were incubated to enable complex formation, which was administered to the wells, ensuring even distribution. The cells were cultured at 37°C with $CO_2$ for 2-4 days prior to conducting live imaging analysis.

Functional Assay: The neuroblastoma cell line SH-SY5Y obtained from ATCC (catalog number: CRL-2266) was used for the functional assay. SH-SY5Y cells were cultured as described elsewhere[55]. pGlo-22 F was transfected in SH-SY5Y cells 24–36 h. The transfected cells were harvested and seeded in 96-well plates at a density of 2 × 104 cells for 24h. On the day of the cAMP assay, the media was removed, and D-luciferin sodium salt (GoldBio, 150 μg/ml) in HEPES (Thermofisher) containing Hanks' Balanced Salt Solution (HBSS) buffer (pH 7.4) was added to the wells and incubated for 30 min at 37 ∘C. After the treatment with forskolin (12.5 μM), graded concentrations (1 nM, 10 nM, 100 nM, 1 μM, 2 μM, 5 μM, 10 μM, and 20 μM ) of dopamine, bromocriptine, compounds 1 and 34. Luminescence signal was measured using Synergy H1 (Bio Tek) at room temperature.

Live-Cell Imaging: Live-cell imaging was carried out in a chamber maintaining 95% air, 5% $CO_2$, and a stable temperature of 37 °C. The HEK293A cells with dopamine receptor sensors in 24-well glass-bottom dishes was utilized for this process. Then the administration of varying ligand concentrations, observing the cells under a 40x objective lens for specified durations. Image capture was executed using a Nikon Ti-E inverted microscope operated by the Nikon Imaging Software-Elements for Advanced Research. Time-lapse of images was captured every 60 seconds at 300 Frames Per Seconds (FPS), with an exposure time of 100 ms using an ND 8 filter. Images were captured live using the Fluorescein Isothiocyanate (FITC) channel (excitation wavelength at 485 nm, emission wavelength at 525 nm) for dLight1.1 and dLight1.5, and the Texas Red Channel (excitation wavelength at 559 nm, emission wavelength at 630 nm) for Rdlight1. The Image J software facilitated the fluorescent intensity calculations, facilitating analysis.

## Data availability

The data that support the findings of this study are available within the main text and its Supplementary Information. Additional datasets can be found on the GitHub repository at https://github.com/temisobodu/dopamine_receptor/tree/dopa_repo

## Code availability

Full for code can be found on the GitHub repository at https://github.com/temisobodu/dopamine_receptor/tree/dopa_repo

## Online content

Code for EQUIBIND can be found on GitHub repository at https://github.com/HannesStark/EQUIBIND. Code for DIFFDOCK can be found on GitHub repository at https://github.com/gcorso/DIFFDOCK.

## Competing Interest

A.Y works for VANTAI. Others authors declare no competing interests.

## Supplementary Figures

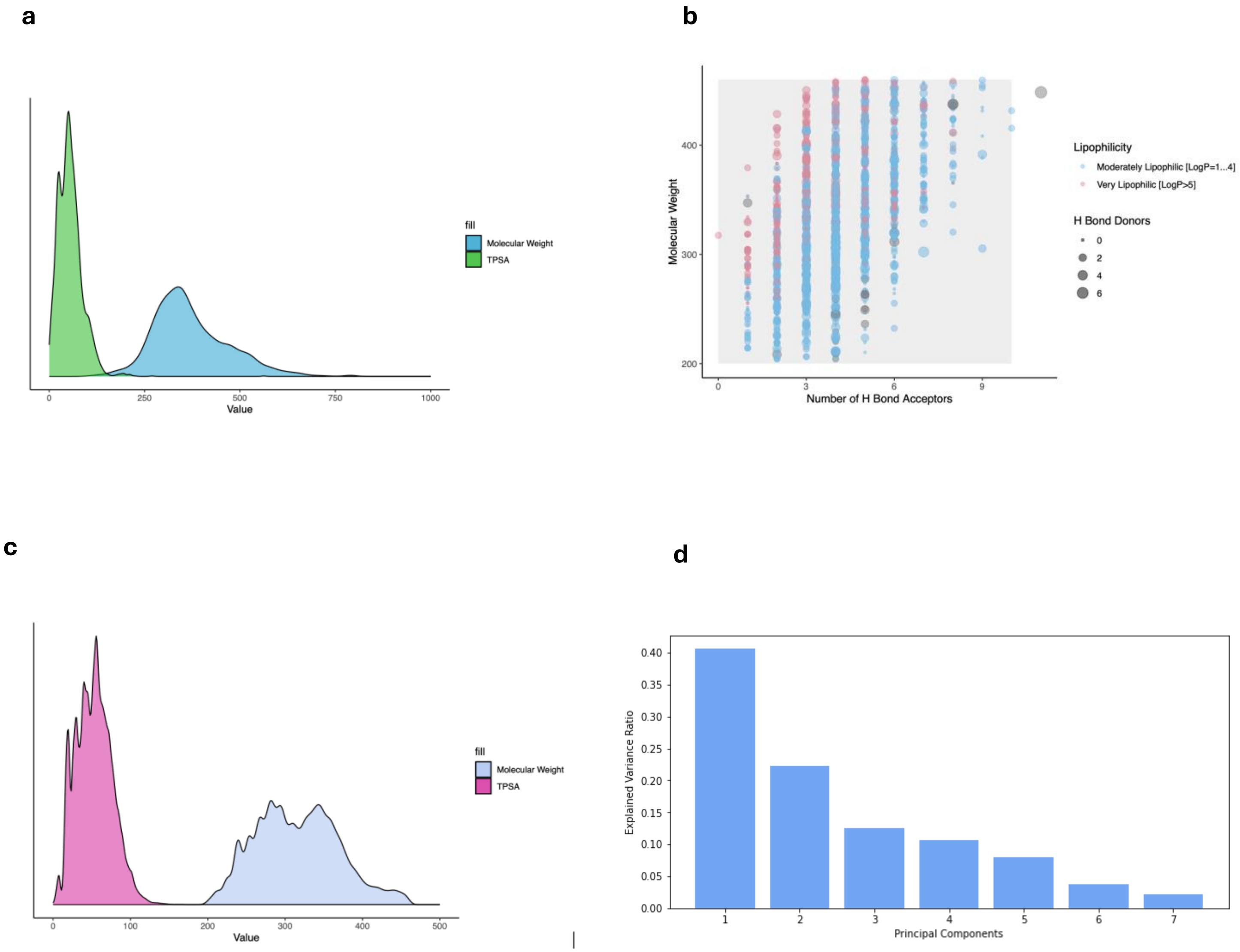

Supplementary Figure 1.a, Density plot depicting molecular weight and Total Polar Surface Area distribution for actives. b, Plot showing the distribution of actives after screening out compounds with HBD > 5, HBA >10 with further categorization based on lipophilicity profile (LogP). Lipinski criteria within the transparent rectangle (n = 1362). c, Density plot depicting the distribution of molecular weight and Total Polar Surface Area (TPSA) for decoys. d, Scree plot visualizing the contribution of each Principal Component to the overall variance.

**a**

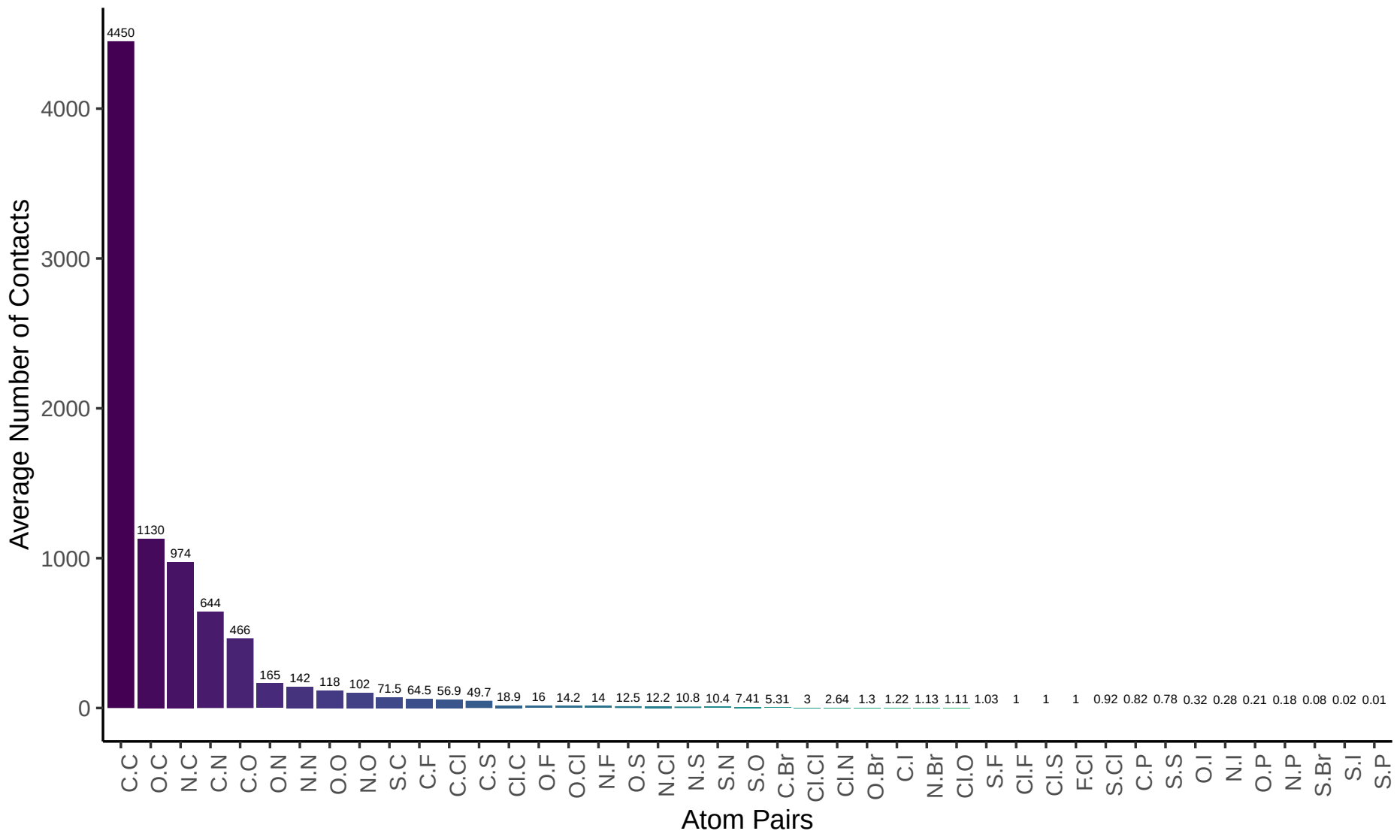

**b**

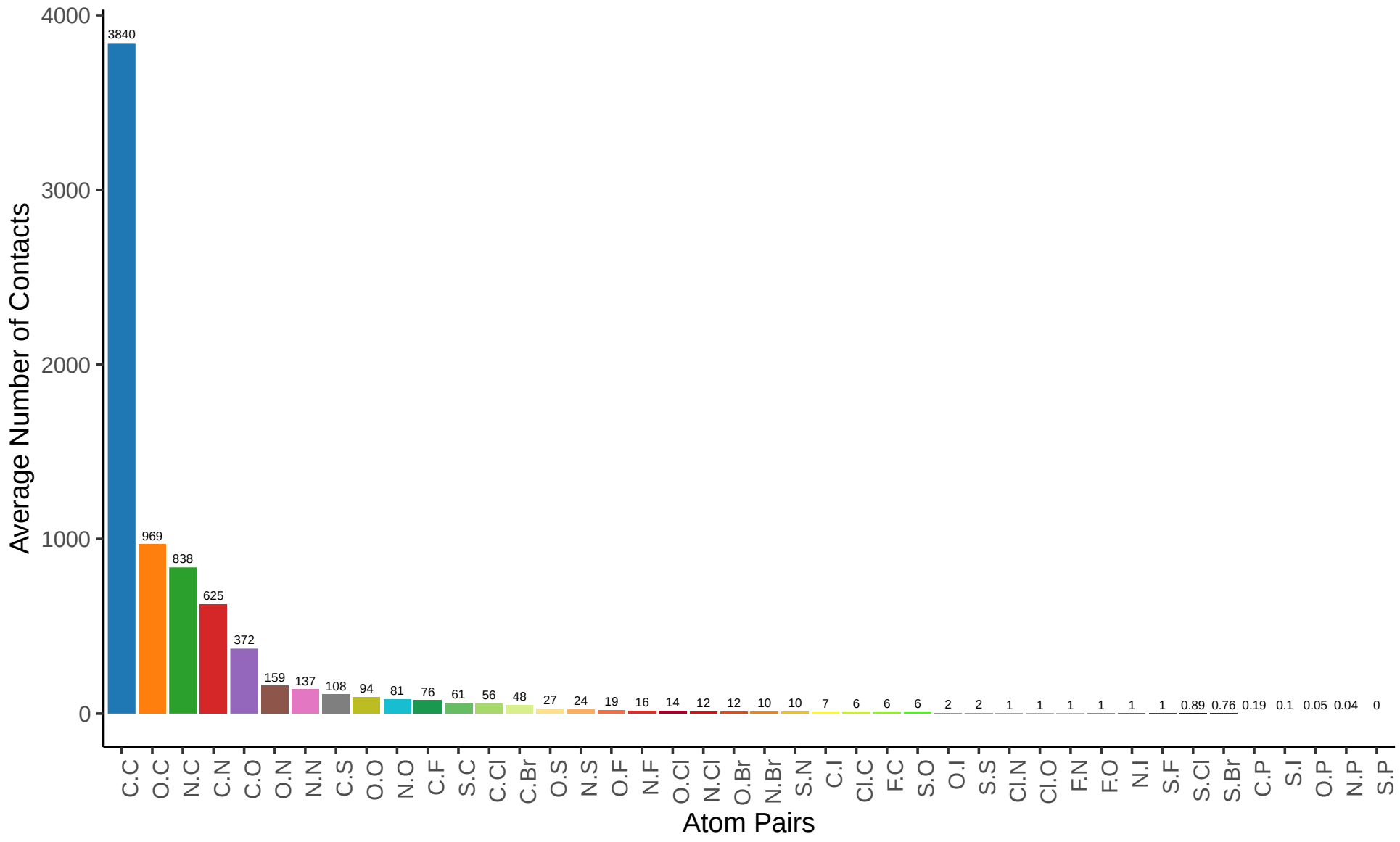

Supplementary Figure 2. **a,** Atomic pair interactions for actives. The chart displays the average number of contacts between protein and ligand atoms within the binding pocket. **b,** Atomic pair interactions for decoys. The chart displays the average number of contacts between protein and ligand atoms within the binding pocket.

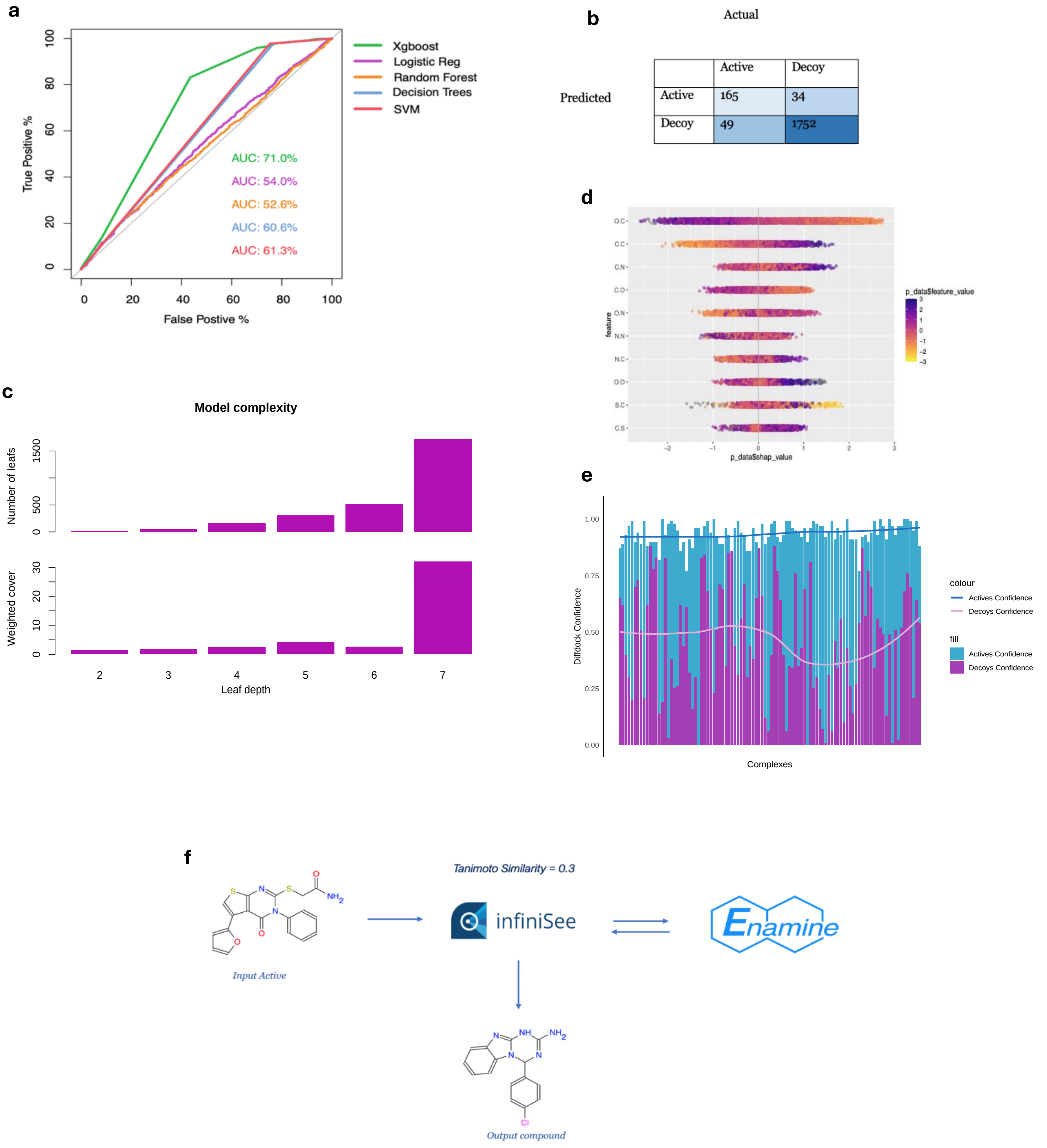

Supplementary Figure 3. **a,** AUROC Curve for Classification Machine Learning Models. **b,** Confusion Matrix for the XGBoost model. **c,** Model complexity relative to tree size and node count. **d,** Graphical representation of SHAP value distribution for weighted features. **e,** Graphical abstract for Ligand-Based Virtual Screening (LBVS) workflow. **f,** DIFFDOCK – normalized confidence benchmark for actives and decoys.

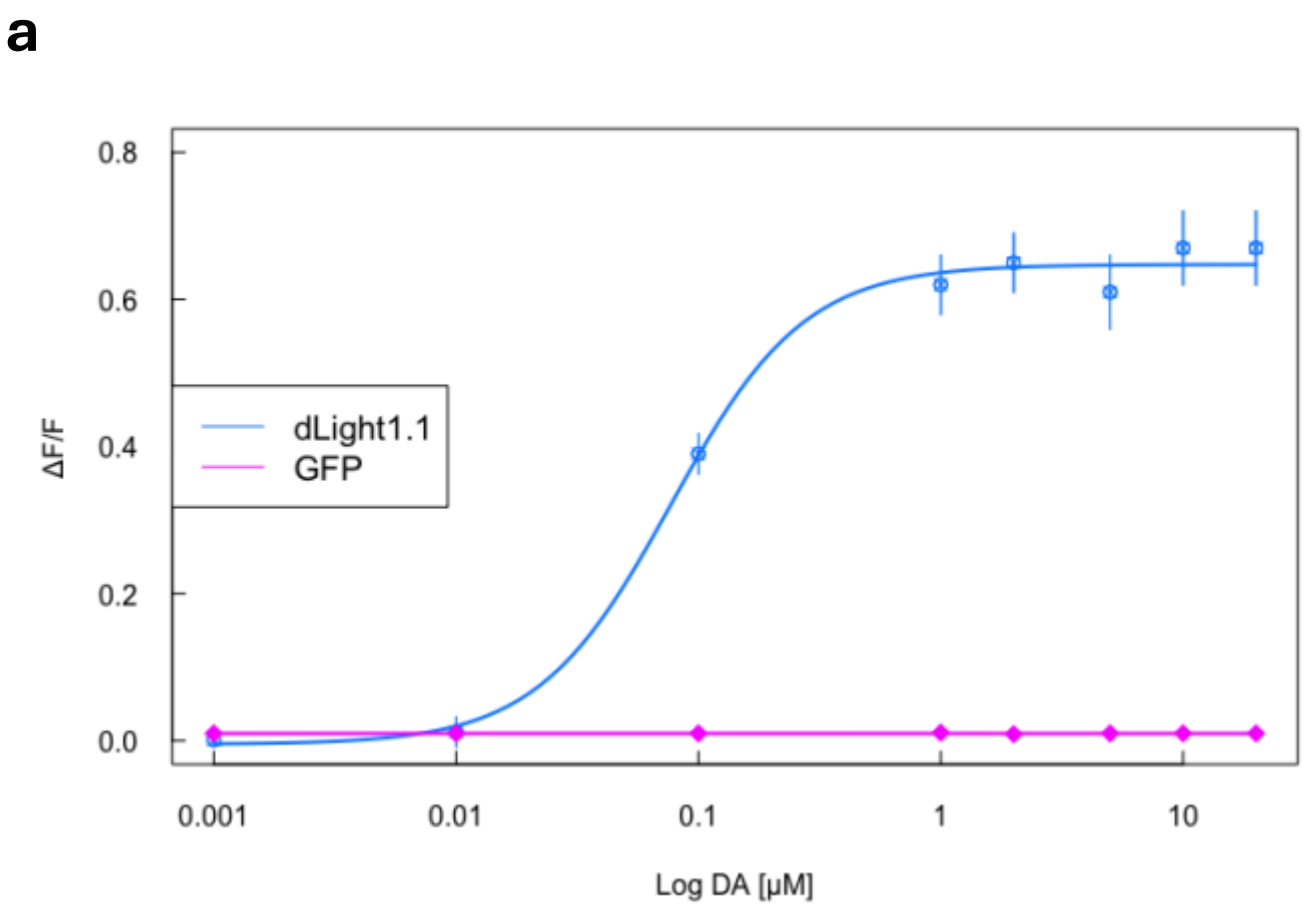

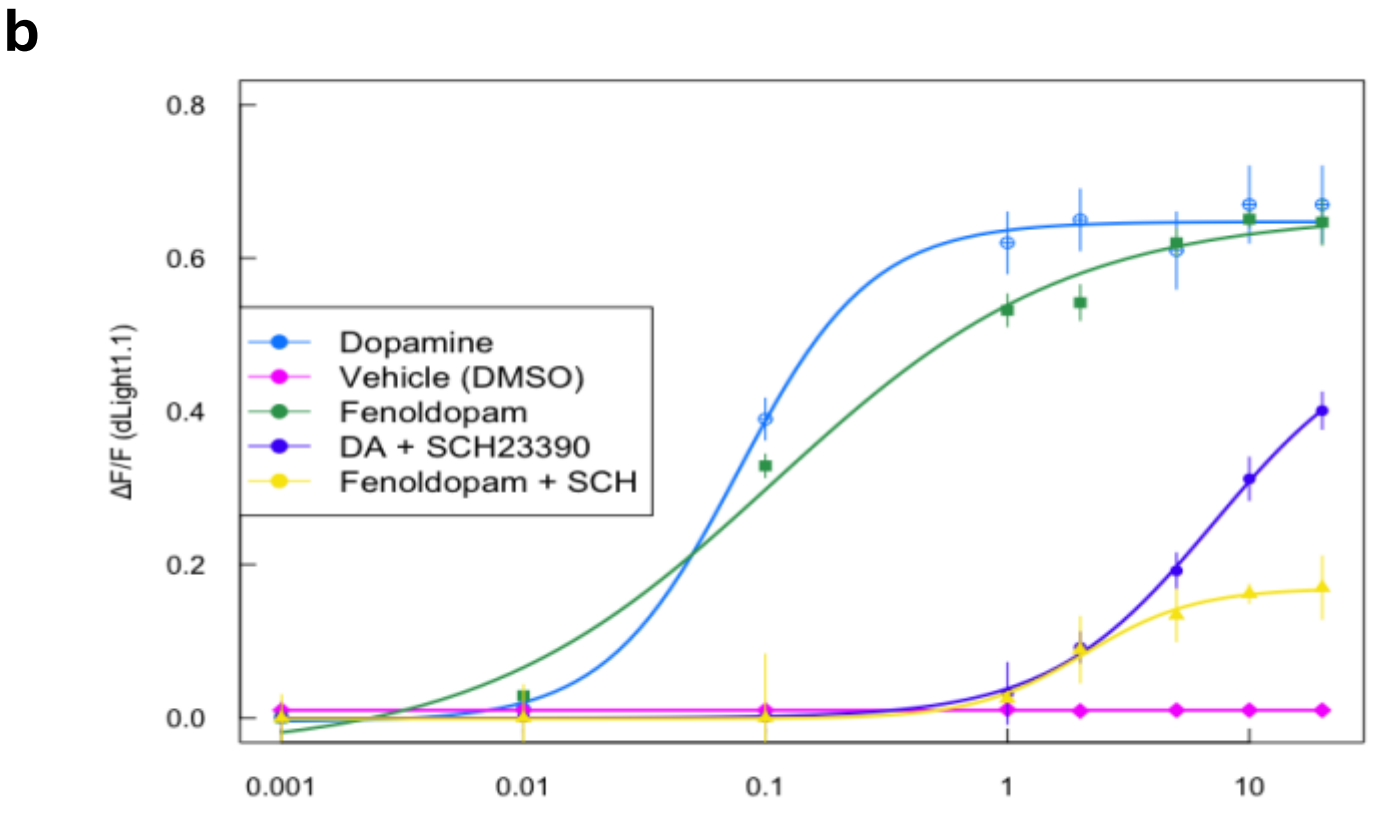

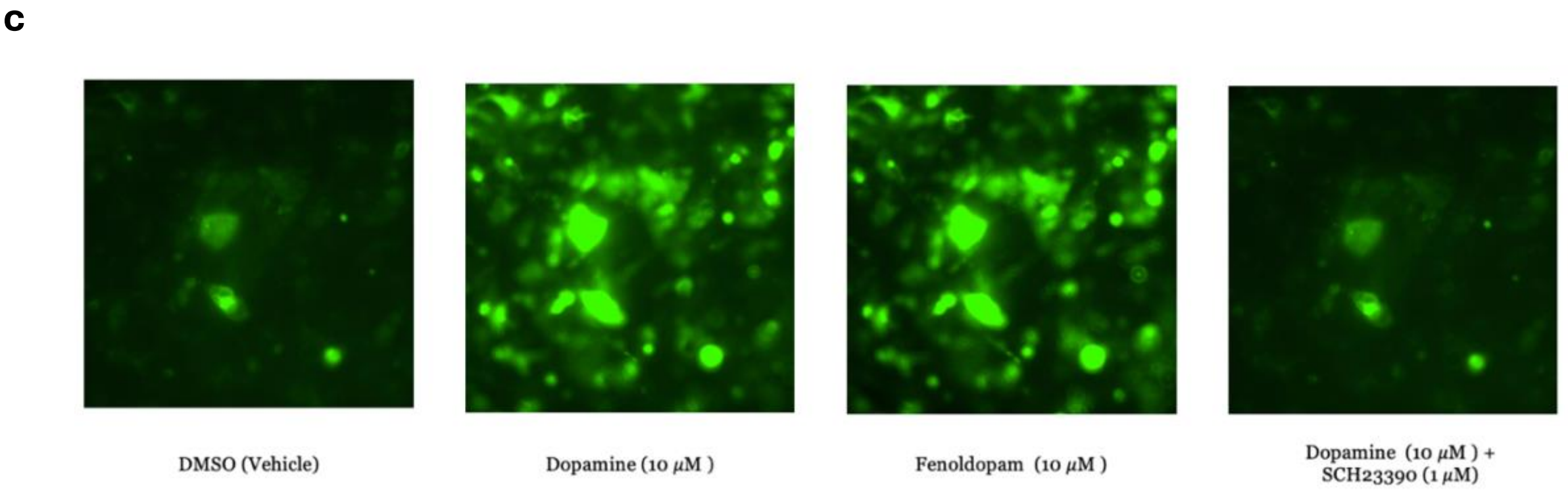

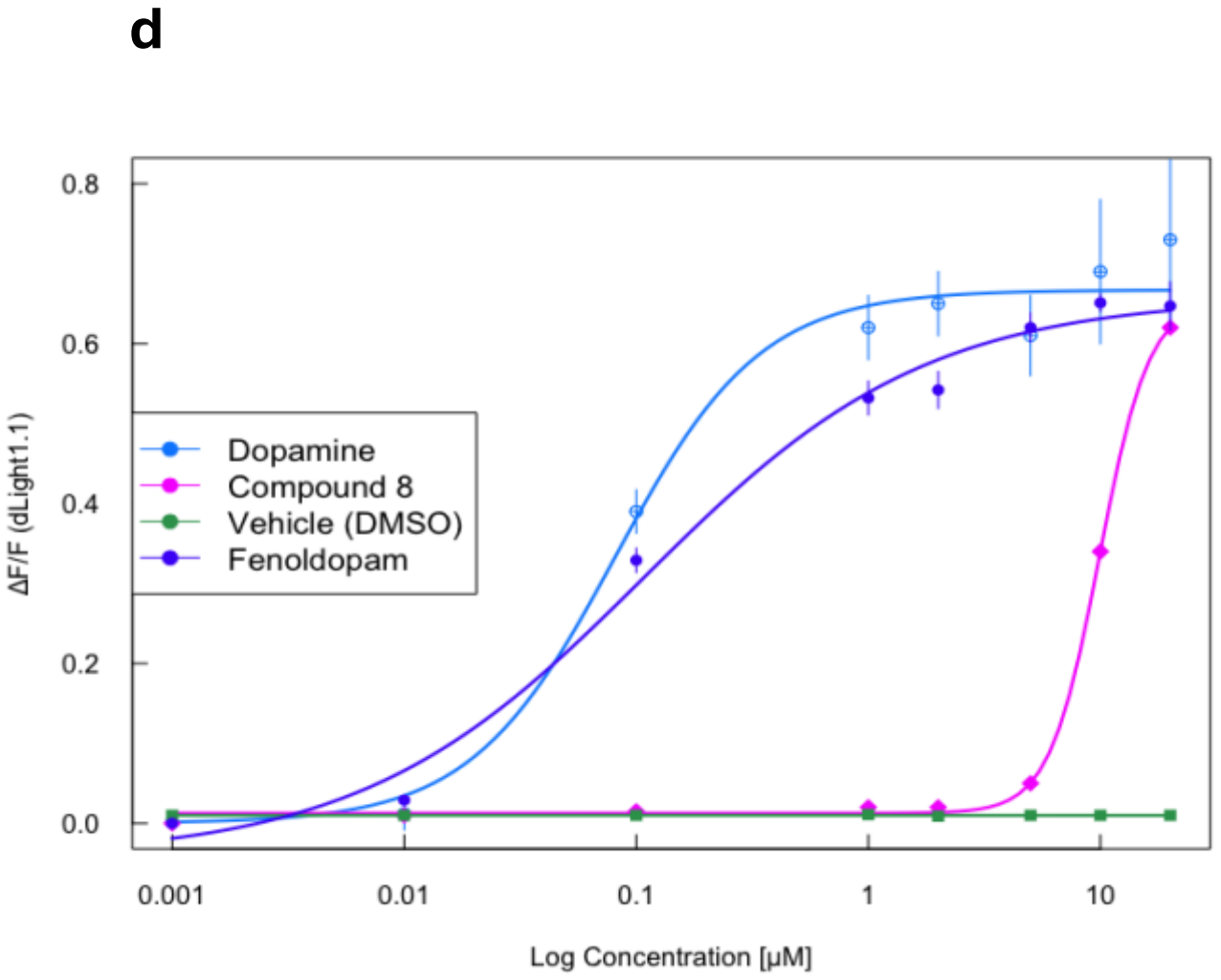

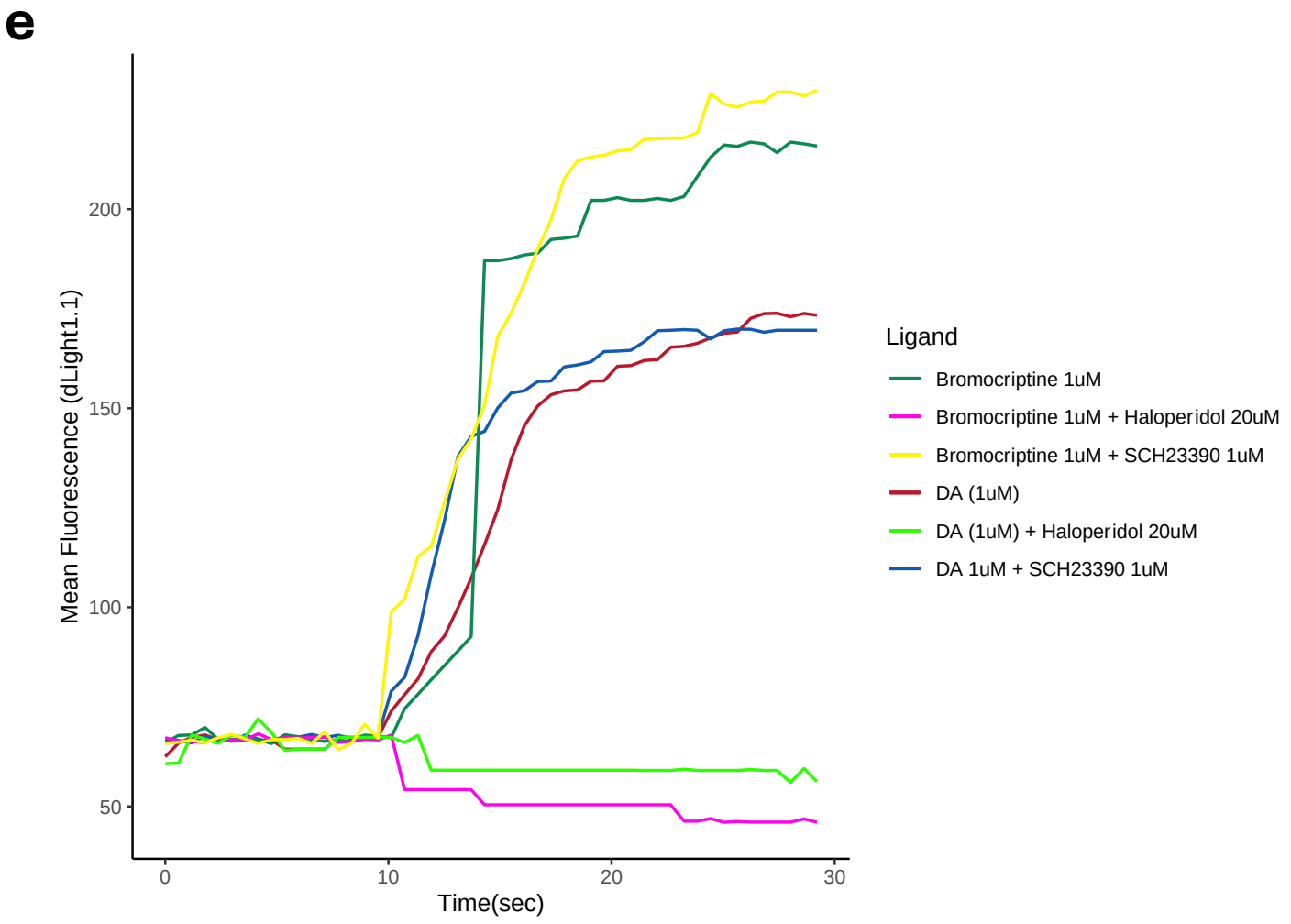

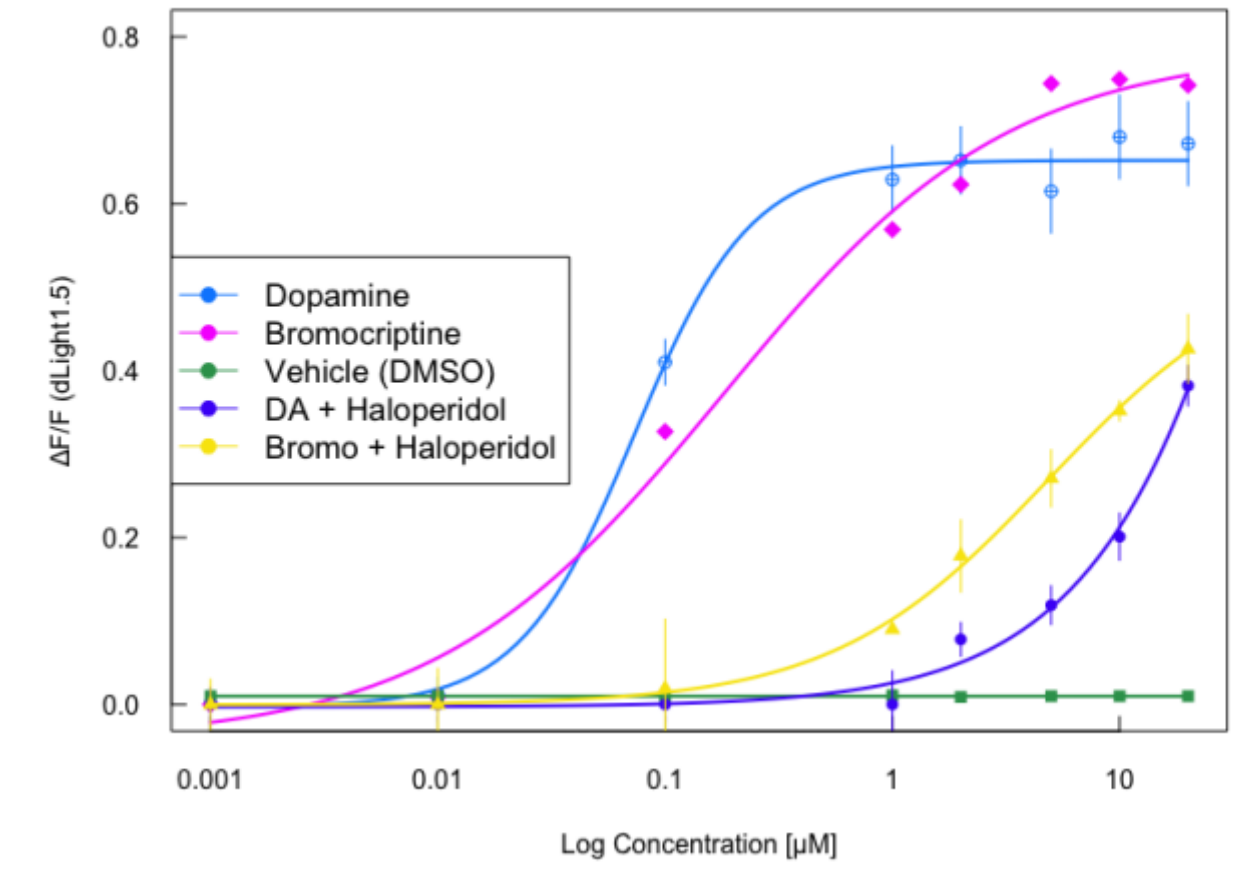

Supplementary Figure 4. **a,** Titration of graded concentrations of dopamine on HEK 293A cells expressing dLight1.1 (mean ± SEM). Fitted with Hill equation (Dopamine's Kd = 0.077± 0.014 $\mu$M). **b,** Titration of graded concentrations fenoldopam and dopamine on HEK 293A cells expressing dLight1.1 with pre-treated SCH2290 (1 $\mu$M) (mean ± SEM). Fitted with Hill equation. **c,** Confocal microscopy visualization shows the activation of dLight1.1 by dopamine (10 $\mu$M) and fenoldopam (10 $\mu$M), the fluorescent response was abolished by SCH23390 (1 $\mu$M). **d,** Titration of graded concentrations Compound 8 (Kd = 9.91± 0.19), dopamine, and fenoldopam on HEK 293A cells expressing dLight1.1 (mean ± SEM). Fitted with Hill equation. **e,** Representative traces of dLight1.5 fluorescence in response to dopamine (1 μM), bromocriptine (1 μM), haloperidol (20 μM), and SCH23390 (1 μM); mean ± SEM, n=5 sessions. **f,** Titration of graded concentrations bromocriptine (Kd = 0.179± 0.069 $\mu$M) and dopamine (Kd = 0.073± 0.012 $\mu$M) on HEK 293A cells expressing dLight1.5 with pre-treated Haloperidol (20 $\mu$M) (mean ± SEM). Fitted with Hill equation.